\documentstyle[12pt]{article}
\newlength{\extraspace}
\newlength{\extraspaces}
\newcounter{fignum}

\newcommand{\be}{\begin{equation}
\addtolength{\abovedisplayskip}{\extraspaces}
\addtolength{\belowdisplayskip}{\extraspaces}
\addtolength{\abovedisplayshortskip}{\extraspace}
\addtolength{\belowdisplayshortskip}{\extraspace}}
\newcommand{\ee}{\end{equation}}

\newcommand{\ba}{\begin{eqnarray}
\addtolength{\abovedisplayskip}{\extraspaces}
\addtolength{\belowdisplayskip}{\extraspaces}
\addtolength{\abovedisplayshortskip}{\extraspace}
\addtolength{\belowdisplayshortskip}{\extraspace}}
\newcommand{\ea}{\end{eqnarray}}

\newcommand{\newsection}[1]{
\vspace{15mm}
\pagebreak[3]
\addtocounter{section}{1}
\setcounter{equation}{0}
\setcounter{subsection}{0}
\setcounter{footnote}{0}
\begin{center}
{\large\sc \thesection. #1}
\end{center}
\nopagebreak
\medskip
\nopagebreak}

\newcommand{\newsubsection}[1]{
\vspace{1cm}
\pagebreak[3]

\addtocounter{subsection}{1}
\noindent{ \sc \thesubsection. #1}
\nopagebreak
\vspace{2mm}
\nopagebreak}

\newcommand{\twomatrixd}[4]{{\left(\begin{array}{cc}\displaystyle #1 &
\displaystyle #2\\[2mm]
\displaystyle #3 & \displaystyle #4 \end{array}\right)}}

\newcommand{\bliep}{\rule[-2mm]{0mm}{7mm}}
\newcommand{\An}{
{\small ${}_1\,\bullet - \bullet - \cdots - \bullet - \bullet_n$}}
\newcommand{\Dn}{\raisebox{-5mm}
{\small ${}_1\;\bliep\bullet - \!
\raisebox{-1mm}{\begin{array}[b]{c} \bliep\bullet \\ | \\ \bullet \end{array}}
\! - \bullet - \cdots - \bullet - \bullet_{n-1}$}}
\newcommand{\Esix}{ \raisebox{-5mm}
{\small\bliep $\bullet - \bullet - \!
\raisebox{-1mm}{\begin{array}[b]{c} \bliep\bullet \\ | \\ \bullet \end{array}}
\! - \bullet - \bullet $}}
\newcommand{\Eseven}{\raisebox{-5mm}
{\small\bliep $\bullet - \bullet -\!
\raisebox{-1mm}{\begin{array}[b]{c} \bliep\bullet \\ | \\ \bullet \end{array}}
\! - \bullet - \bullet - \bullet$}}
\newcommand{\Eeight}{\raisebox{-5mm}
{\small \bliep $\bullet - \bullet -\!
\raisebox{-1mm}{\begin{array}[b]{c} \bliep\bullet \\ | \\ \bullet \end{array}}
\! - \bullet - \bullet - \bullet - \bullet$}}

\newcommand{\ber}{\begin{eqnarray}}
\newcommand{\eer}{\end{eqnarray}}
\newcommand{\ie}{{\it i.e.}}
\newcommand{\cf}{{\it cf}}
\newcommand{\eg}{{\it e.g.\ }}
\renewcommand{\l}{\langle}
\newcommand{\r}{\rangle}
\newcommand{\Bl}{\Bigl\langle}
\newcommand{\Br}{\Bigr\rangle}

\newcommand{\st}{{\tilde{\sigma}}}
\renewcommand{\tt}{{\tilde{t}}}
\newcommand{\dd}[1]{{\partial \over \partial #1}}
\newcommand{\ddt}[1]{{\partial \over \partial t_{#1}}}
\newcommand{\ddtt}[1]{{\partial \over \partial \tt_{#1}}}
\newcommand{\Dd}[2]{{\partial #2 \over \partial #1}}
\newcommand{\Ddt}[2]{{\partial #2 \over \partial t_{#1}}}

\newcommand{\half}{{\textstyle{1\over 2}}}
\renewcommand{\d}{{{\partial}}}

\newcommand{\Z}{{\bf Z}}
\newcommand{\C}{{\bf C}}
\newcommand{\cO}{{\cal O }}
\newcommand{\R}{{\bf R}}
\newcommand{\p}{^{\prime}}

\newcommand{\ra}{\rightarrow}

\newcommand{\inv}{^{-1}}
\newcommand{\HH}{{\cal H}}

\newcommand{\Tr}{{\rm Tr}\,}

\newcommand{\Mgs}{{\cal M}_{g,s}}
\newcommand{\Mgsbar}{\overline{\cal M}_{g,s}}
\newcommand{\Mgo}{{\cal M}_{g,0}}
\newcommand{\Mgobar}{\overline{\cal M}_{g,0}}
\newcommand{\subg}{_{\strut g}}
\newcommand{\subM}{_{\strut M}}
\newcommand{\subt}{_{\strut t}}
\newcommand{\subzero}{_{\strut 0}}
\newcommand{\subone}{_{\strut 1}}
\newcommand{\th}{^{\mit th}}
\newcommand{\res}{{\rm res}\,}

\newcommand{\iss}{& \!\! = \!\! &}
\newcommand{\ppinch}{_{\shortstack{$\scriptstyle g';~ S = X \cup Y $\\
$\scriptstyle i+j=n-2$}}}

\newcommand{\bac}{\begin{array}{c}}
\newcommand{\bacc}{\begin{array}{cc}}
\newcommand{\baccc}{\begin{array}{ccc}}
\newcommand{\barcl}{\begin{array}{rcl}}
\newcommand{\bacccc}{\begin{array}{cccc}}
\newcommand{\baccccc}{\begin{array}{ccccc}}
\newcommand{\baccccccc}{\begin{array}{ccccccc}}
\newcommand{\barclcrcl}{\begin{array}{rclcrcl}}
\newcommand{\bacl}{\begin{array}{cl}}
\newcommand{\bacll}{\begin{array}{cll}}
\newcommand{\eac}{\end{array}}
\newcommand{\fig}{{\it fig.}\ }
\newcounter{tabnum}
\newcommand{\tabel}[2]{
\begin{figure}[t]
\addtocounter{tabnum}{1}
#1
\begin{center}
\parbox{120mm}{\small \bf Table \arabic{tabnum}: \it #2}
\end{center}
\end{figure}}
\newcommand{\nonu}{\nonumber \\[.5mm]}

\newcommand{\Q}{Q}

\newcommand{\dbar}{{\overline{\partial}}}

\newcommand{\bbar}{{\overline{b}}}

\newcommand{\ibar}{{\overline{\imath}}}
\newcommand{\jbar}{{\overline{\jmath}}}

\newcommand{\xbar}{{\overline{x}}}
\newcommand{\zbar}{{\overline{z}}}
\newcommand{\zzbar}{{\bar{z}}}

\newcommand{\qbar}{{\overline{q}}}

\newcommand{\Gbar}{{\overline{G}}}
\newcommand{\Lbar}{{\overline{L}}}
\newcommand{\Qbar}{{\overline{Q}}}
\newcommand{\Tbar}{{\overline{T}}}

\newcommand{\mubar}{{\overline{\mu}}}

\newcommand{\thetabar}{{\overline{\theta}}}
\newcommand{\gammatil}{{\tilde{\gamma}}}
\newcommand{\cc}{{\rm c.c.}}
\newcommand{\Rhat}{\widehat{R}}

\newcommand{\OO}{{\cal O}}

\newcommand{\Int}{\mathop{\int}}
\newcommand{\fr}[2]{{\textstyle {#1 \over #2}}}
\begin{document}
\addtolength{\baselineskip}{.7mm}
\thispagestyle{empty}
\begin{flushright}
{\sc IASSNS-HEP}-91/91\\
December 1991
\end{flushright}
\vspace{.1cm}

\begin{center}
{\Large\sc Intersection Theory, Integrable Hierarchies\\[4mm]
and Topological Field Theory\footnote{Lectures presented at the
Carg\`ese Summer School on {\sl New Symmetry Principles in Quantum Field
Theory}, July 16-27, 1991.}}\\[3cm]
{\sc Robbert Dijkgraaf\footnote{Research supported by the W.M.
Keck Foundation}}\\[3mm]
{\it School of Natural Sciences\\[2mm]
Institute for Advanced Study\\[2mm]
Princeton, NJ 08540}\\[35mm]

{\sc Abstract}
\end{center}
\begin{quote}
\noindent In these lecture notes we review the various relations between
intersection theory on the moduli space of Riemann surfaces, integrable
hierarchies of KdV type, matrix models, and topological field theory. We focus
in particular on the question why matrix integrals of the type considered by
Kontsevich naturally appear as $\tau$-functions of integrable hierarchies
related to topological minimal models.
\end{quote}
\vfill
\newpage

\newsection{Introduction}

The last two years have seen the emergence of a beautiful new subject in
mathematical physics. It manages to combine a most exotic range of disciplines:
two-dimensional quantum field theory, intersection theory on the moduli space
of Riemann surfaces, integrable hierarchies, matrix integrals, random surfaces,
and many more. The common denominator of all these fields is
two-dimensional quantum gravity or, more general, low-dimensional string
theory. Here the application of large-$N$ techniques in matrix models, that are
used to simulate fluctuating triangulated surfaces
\cite{matrix}-\cite{kazakov}, has led to complete solvability
\cite{3}-\cite{c=1}. (See \eg the review papers \cite{matrix-review}, and also
the lectures of S. Shenker in this volume.) Shortly after the onset of the
remarkable developments in matrix models, Edward Witten presented compelling
evidence for a relationship between random surfaces and the algebraic topology
of moduli space \cite{witten,survey}. This proposal involved a particular
quantum field theory, known as topological gravity \cite{topgrav}, whose
properties were further established in \cite{distler,verlinde} and generalized
to the so-called multi-matrix models in \cite{dijkwit}-\cite{milnor}. This
subject can, among many other things, be considered as a fruitful application
of
quantum field theory techniques to a particular problem in pure mathematics,
and as such is a prime example of a much bigger program, also largely due to
Witten, that has been taking shape in recent years. It is impossible to do
fully justice to this subject within the confines of these lecture notes. I
will however make an effort to indicate some of the more startling
interconnections.

I think it is fair to say that one of the most exciting recent developments in
this field has been the work of the Russian mathematician Maxim Kontsevich
\cite{kontsevich1}-\cite{kontsevich3}. He did not only prove rigorously the
first of a set of conjectures of Witten relating the calculus of intersection
numbers on the moduli space of curves to integrable hierarchies of KdV type. In
the process of that, he also derived a concrete `matrix integral
representation' of the string partition function. This is quite a curious
result, since this matrix integral is in no obvious way related to the matrix
models that initiated all these developments. In fact, one could say that
Kontsevich's model is used to triangulate moduli space, whereas the original
models triangulated Riemann surfaces.

Let me first state the main results that will be the revolving point of
these notes. Our starting point will be the particular
two-dimensional quantum field theory know as topological gravity.
Let ${\cal O}_n$ denote the observables in this theory and $t_n$ the coupling
constants to these operators. In the field theory we can consider correlation
functions of the type
\be
\l \cO_{n_1} \cdots \cO_{n_s}\r\subg,
\ee
where $\l\cdots\r_g$ denotes the expectation value on a (connected) surface
with $g$ handles. These correlation functions represent, more or less by
definition, characteristic numbers of the moduli space of Riemann surfaces. The
string partition function $\tau(t)$ is defined as the generating functional of
all possible correlation functions on all possible Riemann surfaces, not
necessarily connected. That is, $\tau(t)$ has an asymptotic expansion of the
form
\be
\tau(t) = \exp \sum_{g=0}^\infty \Bl \exp \sum_n
t_n \cO_n \Br\subg
\label{qft}
\ee
 Witten's conjecture states that, as anticipated in our
notation,
\be
\hbox{$\tau(t)$ \it is a tau-function of the KdV hierarchy.}
\label{tau}
\ee
We will explain the concept of a $\tau$-function in great detail in section 3.
It is a very useful and common notion in the study of integrable hierarchies.
Essentially it implies that from $\tau(t)$ we can construct solutions to the
famous Korteweg-de Vries equation
\be
\Dd t u = u u' + {1\over6}u'''
\ee
and all its generalizations. The relevant point is that $\tau(t)$ is completely
calculable. This integrable structure was first found in the matrix model
description of random surfaces or Euclidean quantum gravity, and so Witten's
conjecture can be restated as the equivalence of quantum and topological
gravity. The intuition behind this remarkable equivalence is some kind of
poorly understood universality in theories of two-dimensional
gravity.\footnote{In the case of two-dimensional {\it gauge} theories the
equivalence of the topological and the non-topological model (also known as
Yang-Mills theory) is much better understood due to recent work of Witten
\cite{topym}.}

The integral representation of $\tau(t)$ that emerges naturally in Kontsevich's
solution involves the integral over a $N\times N$ Hermitian matrix $Y$
of the form
\be
\tau(Z) = \rho(Z)\inv \int dY \cdot \exp \,\Tr [-{1\over2}ZY^2 + {i\over6}Y^3]
\label{kon}
\ee
Here $Z$ is a second $N \times N$ Hermitian matrix, and $\rho(Z)$ is the
one-loop integral
\be
\rho(Z) = \int dY \cdot \exp -{1\over2} \Tr ZY^2
\ee
It is clear that $\tau(Z)$ is conjugation invariant, and so only depends on the
eigenvalues $z_1,\ldots,z_N$ of $Z$. The relation between (\ref{qft}) and
(\ref{kon}) requires a map from the matrix $Z$ to the coupling coefficients
$t_n$. According to Kontsevich this map is given by
\be
t_n = -{1\over n} \Tr Z^{-n}.
\ee
As we will see, the matrix integral (\ref{kon}) has a natural expansion in
Feynman diagrams and each diagram represents the integral over a particular
cell in the moduli space of Riemann surfaces. It can therefore be considered
as a finite-dimensional reduction of the path-integral of topological gravity.

Finally, there is a third and equivalent way to characterize $\tau(t)$, that
naturally arises in topological field theory and in considerations of so-called
loop operators. In this formulation a series of differential operators $L_n$
--- quadratic in the parameters $t_k$ and their derivatives $\d/\d t_k$ ---
annihilate the string partition function \cite{loop,kawai}
\be
L_n \cdot \tau = 0,\qquad\quad n\geq -1.
\label{Ln}
\ee
These operators, that will be consider in greater detail in sections 4 and 7,
form part of a Virasoro algebra---they satisfy the commutation relations
\be
[L_n,L_m]=(n-m) L_{n+m}.
\ee
The constraints (\ref{Ln}) can be used to derive recursion relations for the
correlation functions and thus provide an alternative way to arrive at the
asymptotic expansion for $\tau(t)$. From a computational perspective,
(\ref{Ln}) is perhaps the most practical characterization of $\tau(t)$.

There exist a whole hierarchy of generalizations of all these structures, where
the results become more and more incomplete and conjectural as we move away
from the original setting. The most simple generalization involves a choice of
integer $p\geq 2$, with the case $p=2$ corresponding to intersection theory on
moduli space. In the case there are well-defined candidates for the
corresponding problem in algebraic geometry, the integrable hierarchy, and the
matrix integral. As we will see this can naturally be considered the $A_{p-1}$
case in a series of models labeled by a simply-laced Lie group $G$, \ie, $G$ is
of type $A_n$, $D_n$, $E_6$, $E_7$, or $E_8$. The most general case is based on
an arbitrary two-dimensional topological field theory, and includes, among
others, sigma-models with an almost complex target space. This field is known
as
{\it topological string theory} and its status as an integrable system is very
unclear, to say the least.

Let me close this introduction with a short exposition of how this all is
related to $c<1$ string theory, where one studies minimal conformal field
theories on fluctuating surfaces. These minimal models are labeled by two
(relative prime) integers $p$ and $q$---the labeling $(p,q)$ is symmetric in
$p$ and $q$. The models can be considered as isolated points in an ill-defined
space of all two-dimensional quantum field theories coupled to gravity, as is
symbolically illustrated in \fig 1. A neighbourhood of a minimal model consists
of massive QFT's obtained by perturbation. The counting of physical observables
is rather complicated \cite{lian}, but there is a simple subsector---which does
not involve the reparametrization ghosts---and in this subsector we find
\be
\hbox{$\#$ states} = \half (p-1)(q-1).
\ee
This finite number of states is the defining property of minimal models.
Through the work of Kazakov \cite{kazakov}, one can use matrix models to
represent both the random surfaces and the conformal field theory by choosing
the right critical behavior of the matrix potential. This leads to a uniform
description of all models with identical $p$, where a chain of $(p-1)$ matrices
suffices. The second number $q$ now labels the order of
criticality. All models with fixed $p$ are related through the $p\th$ KdV
hierarchy \cite{douglas}, which moves horizontally in \fig 1. That is, their
partition functions are given by the {\it same} $\tau$-function, evaluated at
different values of the coupling coefficients $t_n$.
\begin{figure}[t]
\addtocounter{fignum}{1}
\begin{center}
\mbox{
\setlength{\unitlength}{.7mm}
\begin{picture}(80,50)(0,0)
\put(20,20){\makebox(0,0){$\circ$}}
\put(30,20){\makebox(0,0){$\bullet$}}
\put(40,20){\makebox(0,0){$\bullet$}}
\put(50,20){\makebox(0,0){$\bullet$}}
\put(20,30){\makebox(0,0){$\circ$}}
\put(30,30){\makebox(0,0){$\bullet$}}
\put(40,30){\makebox(0,0){$\bullet$}}
\put(50,30){\makebox(0,0){$\bullet$}}
\put(20,40){\makebox(0,0){$\circ$}}
\put(30,40){\makebox(0,0){$\bullet$}}
\put(40,40){\makebox(0,0){$\bullet$}}
\put(50,40){\makebox(0,0){$\bullet$}}
\put(20,50){\makebox(0,0){$\circ$}}
\put(30,50){\makebox(0,0){$\bullet$}}
\put(40,50){\makebox(0,0){$\bullet$}}
\put(50,50){\makebox(0,0){$\bullet$}}
\put(60,20){\makebox(0,0){$\cdots$}}
\put(60,30){\makebox(0,0){$\cdots$}}
\put(60,40){\makebox(0,0){$\cdots$}}
\put(60,50){\makebox(0,0){$\cdots$}}
\put(20,60){\makebox(0,0){$\vdots$}}
\put(30,60){\makebox(0,0){$\vdots$}}
\put(40,60){\makebox(0,0){$\vdots$}}
\put(50,60){\makebox(0,0){$\vdots$}}
\put(10,30){\vector(0,1){20}}
\put(30,10){\vector(1,0){20}}
\put(40,5){\makebox(0,0){$q$}}
\put(5,40){\makebox(0,0){$p$}}
\put(20,14){\makebox(0,0){{\small $1$}}}
\put(14,20){\makebox(0,0){{\small $2$}}}
\end{picture}}\\[5mm]
\parbox{120mm}{\small \bf Fig.\ \arabic{fignum}: \it The space of $c<1$ field
theories coupled to quantum gravity contains as special point the $(p,q)$
minimal models. The models of type $(p,1)$, here indicated by open dots, should
be considered as topological field theories, and are the main subject of these
notes.}
\end{center}
\end{figure}

Models of type $(p,1)$ are strictly speaking not well-defined conformal field
theories---according to the above formula they have zero physical fields.
However, one can make sense out of these models as topological field theories,
the so-called {\it topological minimal models}, and they will form the main
topic of these lecture notes. Since the orbit of a topological model under the
KdV flows will cover all other $(p,q)$ models with fixed $p$, this is in fact a
very elegant and powerful way to study the general case. We will see that the
behavior of many quantities simplifies dramatically at the topological point
$q=1$, in particular the so-called Baker-Akhiezer function is exactly
calculable, and this will lead to an explicit solution. As mentioned before
this leads us again, through the work of Kontsevich, to matrix models, although
of a very different kind. In fact, Kontsevich's generalized matrix model
interpolates in a natural way between the models with different $p$---its
allows `vertical' deformations in \fig 1---and can thus be seen as
complementary to the original double-scaled matrix models.

In the following sections we will try to explain the various aspects of these
beautiful but at first sight highly mysterious results and their many
generalizations. We will focus in particular on the question why matrix
integral
   s
arise naturally in integrable hierarchies, a point that is also addressed in a
number of recent publications \cite{kontsevich3,morozov,i-z}. These notes are
organized as follows: We start our discussion with (generalized) intersection
theory on moduli space in \S2. The integrable hierarchies of KdV type that
emerg
   e
in that context are described in \S3, which follows very closely the exposition
   of
Segal and Wilson \cite{segal-wilson}. In \S4 we show how the apparatus of
Grassmannians can be applied to our specific problem and how all this is
related
to matrix integrals. The quantum field theories that have led us to these
result
   s
and conjectures first enter in \S5. Here we only consider the general
properties
of topological field theories that obey factorization conditions. This is made
much more explicit in \S6, where various two-dimensional models are
constructed,
with particular emphasis on the so-called Landau-Ginzburg models. In \S7 we
finally discuss topological gravity, and more generally topological string
theor
   y.
Since we covered much of this material on a previous occasion \cite{notes}, we
will be here far less complete. In particular, the Virasoro constraints and
thei
   r
relation with loop equations will only be touched upon. Finally a point of
caution: we have not always been extremely careful in our calculations with all
`irrelevant' numerical constants. We refer the reader to the literature for the
factors of $i$ and $2\pi$.

\newsection{Intersection Theory}

We start from the quantum field theory end, or more precisely from the finite
dimensional cohomology problem to which the quantum field theory, by
construction, can be reduced to. We will return to the description of the full
quantum field theories in the sections 5--7.

\newsubsection{The moduli space of curves}

Let us briefly recall some well-known facts about the moduli space of smooth
complex curves or Riemann surfaces. A given topological surface $\Sigma$ with
$g$ handles and $s$ marked points $x_1,\ldots,x_s$ can be made into a complex
manifold by endowing it with a complex structure $J$, that is, a local linear
map on the tangent bundle that satisfies $J^2=-1$ and the integrability
condition $DJ=0$. Two complex structures are considered equivalent if they are
related by a diffeomorphism. This leaves us actually with a finite-dimensional
space of inequivalent complex structures on the surface, known as the moduli
space $\Mgs$. By Riemann-Roch this is a space of complex dimension
\be
\dim \Mgs = 3g-3+s.
\ee
(The moduli spaces ${\cal M}_{0,s}$ are only well-defined for $s\geq3$, the
exceptional case ${\cal M}_{1,0}$ equals ${\cal M}_{1,1}$ and is thus
one-dimensional.) Moduli spaces are not everywhere smooth, since surfaces with
accidental higher symmetry give rise to so-called orbifold points.

One way to think about the complex structure on $\Sigma$ is as the conformal
class of a metric $h_{\mu\nu}$. Indeed, a metric defines a complex structure
through
\be
J_{\mu}{}^\nu = \sqrt{h}\, \epsilon_{\mu\lambda} h^{\lambda\nu},
\ee
with $\epsilon_{\mu\nu}$ the Levi-Civita symbol, and it is a useful fact that
all $J$'s can be obtained in this way. Since the definition of $J$ is
independent of local rescalings of $h_{\mu\nu}$, we can represent $\Mgs$
alternatively as the space of metrics modulo local rescalings and
diffeomorphisms.

The moduli space $\Mgs$ has a boundary and there exists a direct, intuitive
interpretation of the points at the boundary --- they represent degenerate
surfaces. There are basically two ways in which a surface can degenerate. If
we think in terms of a conformal class of metrics, the surface can either form
a node---or, equivalently, a long neck---or two marked points can collide. The
boundary of $\Mgs$ can be thought to lie at infinity. One would like to
compactify the moduli space by adding points at infinity, not unlike how one
compactifies the plane $\R^n$ to the $n$-dimensional sphere. In this case the
points at infinity represent particular Riemann surfaces. The Deligne-Mumford
or `stable' compactification \cite{deligne-mumford} adds three types of these
points, as is illustrated in \fig 2.

{\it (i)} The process in which two points $x_1$ and $x_2$ `collide' if
$q=x_1-x_2$ tends to zero can (after a coordinate transformation $x \ra x/q$)
alternatively be described as the process in which a sphere, that contains
$x_1$ and $x_2$ at fixed distance, pinches off the surface by forming a neck of
length $\log q$. These two descriptions are fully equivalent, but the latter is
actually more in the spirit of conformal field theory, since we see in an
obvious way the operator product expansion emerge. It is also suggestive of
another end point configuration. Here the natural final configuration is not
simply the surface with $x_1=x_2$, but it consists of a separate sphere
containing the points $x_1$, $x_2$ and a third point where the infinite long
tube was attached, together with the original surface with one marked point
less. In the stable compactification we add this configuration as limit point,
which symbolically can be written as the process
\be
(g,s) \ra (g,s-1) + (0,3).
\ee
Recall that the thrice-punctured sphere has a unique complex structure, so
there are no moduli associated to it and this boundary component has
codimension one. The crucial property of this compactification is that the
points $x_i$ never come together. The two other degeneration modes are more
straightforward.

{\it (ii)} If a cycle of non-trivial homology pinches, we replace the surface
by a surface with one handle less and two extra marked points---the attachment
points of the infinitely thin handle,
\be
(g,s) \ra (g-1,s+2).
\ee

{\it (iii)} In a similar spirit: in case a dividing cycle pinches, the
resulting surface consists of two disconnected surfaces of genus $g'$ and
$g-g'$, each having one extra puncture
\be
(g,s) \ra (g,s'+1)+ (g-g',s-s'+1).
\ee

It can be shown that this prescription makes $\Mgs$ into a compact smooth
(orbifold) space $\Mgsbar$. We now want to consider its cohomology ring
$H^*(\Mgsbar)$. A particular set of elements has been considered by Mumford,
Morita, and Miller \cite{mumford}--\cite{miller}. These classes are constructed
as follows \cite{witten}. There exist $s$ natural line bundles $L_1,\ldots,L_s$
on the moduli space. The fiber of the bundle $L_i$ at a point $\Sigma \in \Mgs$
is the cotangent space to the point $x_i$ on the surface $\Sigma$. These line
bundles have first Chern classes that in de Rham cohomology are represented by
the curvature $F_i$ of an arbitrary $U(1)$ connection on $L_i$. This defines
for us the $2n$-dimensional classes
\be
\sigma_n(i) = c_1(L_i)^n \in H^{2n}(\Mgs), \quad \sigma_n(i)=
\underbrace{F_i \wedge \ldots \wedge F_i}_n.
\ee
It is a non-trivial property that these are stable classes, \ie, that their
definition actually extends to the {\it compactified} moduli space $\Mgsbar$.
The reason is basically that the line bundles $L_i$ remain well-defined in the
stable compactification, since the points $x_i$ will never collide. At
infinity they remain at finite distance while being separated from the bulk of
the surface on a `frozen' 3-punctured sphere, which carries no moduli.

We can now define `correlation functions' by pairing the wedge product of the
$\sigma_n$'s with the fundamental class of the moduli space
\be
\l \sigma_{n_1} \cdots \sigma_{n_s} \r\subg \equiv
\l \sigma_{n_1}(1) \cdots \sigma_{n_s}(s),\Mgsbar \r
\ee
Of course, these intersection numbers are only non-zero if the degree of the
total class equals the dimension of moduli space, which gives the `charge
conservation' condition
\be
\sum_{i=1}^s (n_i-1) = 3g-3.
\label{charge}
\ee
Although the Chern classes $c_1(L_i)$ are in principle defined in integer
cohomology, their intersection numbers will be in general rational numbers, due
to the orbifold nature of the moduli space.

The amazing result, first conjectured by Witten \cite{witten,survey} on the
basis of the matrix model results and later proved rigorously by Kontsevich
\cite{kontsevich1}, states that all these intersection numbers are explicitly
computable because their generating function obeys the equations of the KdV
hierarchy. To state this result more precisely, it is convenient to rescale
and relabel the classes $\sigma_n$ as
\be
\cO_{2n+1} = (2n+1)!!\cdot\sigma_n,\qquad n \geq 0,
\ee
After this substitution the theorem states that the string partition function
\be
\tau(t) = \exp \sum_{g=0}^\infty \Bl \exp\sum_{k\ \rm odd} t_k \cO_k \Br\subg
\ee
is a $\tau$-function of the KdV hierarchy. In this sense intersection theory on
the (universal) moduli space is completely integrable. We note that due to the
charge conservation condition (\ref{charge}), it is very easy to introduce a
string coupling constant $\lambda$ that keeps track of the genus $g$ of the
different surfaces that appear in the partition function $\tau(t)$.
Indeed, if we rescale the coupling constants by
\be
t_k \ra \lambda^{{k\over3}-1}t_k,
\ee
we find
\be
\tau(t) \ra \exp \sum_{g=0}^\infty \lambda^{2g-2}
\Bl \exp\sum_{k\ \rm odd} t_k \cO_k \Br\subg
\ee
Before we discuss the proof of the integrability of $\tau(t)$, let us first
discuss some weaker results related to the `trivial' nature of the classes
$\sigma_0$ and $\sigma_1$.

\newsubsection{The puncture and dilaton equations}

The only consequence of the insertion of the `puncture operator' $\sigma_0$ in
a correlation function is that the same cohomology class is now integrated over
$\overline{\cal M}_{g,s+1}$ instead of $\Mgsbar$. Since there is a natural
projection map $\pi: \overline{\cal M}_{g,s+1} \ra \Mgsbar$, where one simply
forgets the position of the extra point, one might tend to conclude that the
classes can be simply pull-backed and that thus the correlation function
including $\sigma_0$'s vanish by dimensional reasons. This is actually not
true, due to subtleties at the divisor at infinity, where the forgotten point
comes close to the other positions. This leads to extra corrections that can be
expressed through the so-called puncture equation, due to Deligne
\cite{deligne}, which is explained in more detail in \cite{dijkwit},
\be
\l \sigma_0 \cdot \sigma_{n_1} \cdots \sigma_{n_s} \r\subg =
\sum_{i=1,\; n_i\neq 0}^s \l \sigma_{n_1} \cdots \sigma_{n_i-1} \cdots
\sigma_{n_s} \r\subg
\label{puncture}
\ee
This relation can be used to eliminate all $\sigma_0$'s from a particular
correlation function. It is of course only valid if both sides of the equation
are well-defined, which is not the case in genus zero with less than three
insertions. In fact, there we have the additional relation
\be
\l \sigma_0 \sigma_0 \sigma_0 \r\subzero = 1,
\label{sphere}
\ee
a simple consequence of ${\cal M}_{0,3} =$\ point.

A similar result can be obtained for the insertions of the next operator in
line---the `dilaton operator' $\sigma_1$. Since $\sigma_1$ is the first Chern
class $c_1(L)$ it can be shown to calculate the degree of the canonical line
bundle of a genus $g$ surface with $s$ punctures \cite{survey}. The dilaton
equation exactly expresses this result
\be
\l \sigma_1 \cdot \sigma_{n_1} \cdots \sigma_{n_s} \r\subg =
(2g-2+s) \l \sigma_{n_1} \cdots \sigma_{n_s} \r\subg
\label{dilaton}
\ee
Here again we must be careful if the right-hand side is ill-defined. This
happens in genus one with no punctures. Here one finds
\be
\l \sigma_1 \r\subone = {1\over 24}.
\label{torus}
\ee

The above equations allow one to eliminate all operators $\sigma_0$ and
$\sigma_1$. Actually this, together with the selection rule (\ref{charge}),
suffices to reduce any intersection number in genus zero or one to
(\ref{sphere}) and (\ref{torus}) respectively. On the sphere the final
result reads
\be
\l \sigma_{n_1} \cdots \sigma_{n_s} \r\subzero = {(n_1+\ldots+n_s)! \over
n_1!\ldots n_s!}
\ee
On the torus it is more difficult to give a closed form expression, but one way
to state the result is by means of generating functions
\be
\Bl \exp \sum_n t_n \sigma_n \Bigr \r\subone
= {1\over 24} \log \, \Bl \sigma_0^3 \cdot \exp \sum_n t_n \sigma_n \Br\subzero
\ee
In this fashion one can verify the KdV relations by hand at low genus.
Similar verification have been done by Witten \cite{survey} in genus two and
Horne \cite{horne} in genus three, using results of Mumford \cite{mumford} and
Faber \cite{faber} respectively.

\newsubsection{Kontsevich's calculation}

There exists a very explicit realization of the moduli spaces $\Mgs$ for $s>0$
due to the work of Strebel \cite{strebel}, Penner \cite{penner}, and others,
that naturally arises in open string field theory. Recall that we can think of
complex structures in terms of conformal classes of metrics. In open string
field theory Riemann surfaces are built out of flat strips (propagators) with
fixed width and flexible lengths $\ell_a$ by glueing them together with a
three-point vertex, in such a way that all the curvature is localized in the
mid-point of the vertex. With $3k$ propagators and $2k$ vertices one produce a
surface of Euler number $\chi=2-2g-s=-k$. The corresponding Feynman diagram is
called a ribbon or fat graph. Some examples are sketched in \fig 3. These
surface have by construction at least one boundary. We can associate a closed
surface with punctures to a ribbon graph $\Gamma$ by glueing onto the boundary
components $C_1,\ldots,C_s$ infinitely long cylinders---which are conformally
equivalent to punctured disks. We have $6g-6+3s$ real variables $\ell_a$
parametrizing the surface, $s$ of which correspond to the total lengths of the
boundaries while the other variables parametrize a top-dimensional simplex in
the moduli space $\Mgs$. It is a powerful result that we can actually obtain
any point of the moduli space once and only once by considering all connected
graphs $\Gamma$ of the appropriate Euler number.

We can define a `path-integral' whose expansion in Feynman diagrams naturally
produces the graphs $\Gamma$. The fundamental idea, basically due to 't Hooft
\cite{thooft}, is to use an Hermitian $N \times N$ matrix $Y$ to represent the
double line propagator, appropriate for a ribbon graph. Indeed, the matrix
integral
\be
Z = \int dY \cdot \exp {1\over\lambda}
\Tr[-{1\over2}Y^2+{1\over3}Y^3]
\label{count}
\ee
has an expansion
\be
\log Z = \sum_\Gamma {1\over n(\Gamma)} \lambda^{2g-2+s} N^s
\ee
where we sum over all connected graphs $\Gamma$; $n(\Gamma)$ is the order of
the automorphism group of the graph $\Gamma$, $\lambda$ is the
string coupling constant. $Z$ simply counts all top-dimensional cells exactly
once, and thus represents an `integral' over moduli space.

Kontsevich has used this `open string field theory' description of moduli space
to calculate the intersection numbers of the classes $\sigma_n$. The first step
is to write the Chern classes $c_1(L_i)$ in terms of the lengths $\ell_a$'s.
Let us concentrate on one boundary component $C_i$ of the surface.
It has total length
\be
p_i = \sum_{a\in I_i} \ell_{a},
\ee
where $I_i$ labels all propagators that contribute to the boundary $C_i$.
According to Kontsevich the subdivision of $p_i$ in the $\ell_a$ can be used to
define a natural connection on the circle bundle, or actually the polygon
bundle, $L_i$ associate to the boundary. The first Chern class of this line
bundle $L_i$ is the given by the curvature of this connection, and takes the
form
\be
c_1(L_i) = \sum_{a,b\in I_i} d\theta_a \wedge d \theta_b,\qquad
\theta_a=\ell_a/p_i.
\ee
We now want to perform the integral of these Chern classes over moduli space.
We do not have the space here to reproduce the complete calculation, which has
many deep subtleties. Let us briefly indicate the crucial steps. First, one
considers not the classes $\sigma_n$ but the `loop operators'
\be
w(z) =\int dp \;e^{-zp}\exp \half p^2 c_1(L)=
\sum_{k\ \rm odd} {1\over k} z^{-k} \cO_k.
\label{wz}
\ee
One then evaluates the correlation function of these operators $w(z)$ for
fixed genus $g$. This is an integral over both $\Mgs$ and the boundary lengths
$p_1,\ldots,p_s$. According to Kontsevich this integral becomes elementary
when expressed in terms of the coordinates~$\ell_a$. With the two-form $\Omega$
defined by
\be
\Omega = \sum_{i=1}^s \half p_i^2 c_1(L_i),
\ee
we see that the integral contains indeed the simple measure factor
\be
e^\Omega \cdot dp_1\wedge \ldots\wedge dp_s = c \cdot \prod_a d\ell_a.
\ee
Surprisingly the constant $c$ does not depend on the graph $\Gamma$. One simply
finds $c=2^k$ for a surface with Euler characteristic $\chi=-k$ (see Appendix C
of \cite{kontsevich3}), and the calculation reduces to
\be
\l w(z_1)\cdots w(z_s) \r\subg = \sum_\Gamma {2^k\over n(\Gamma)} \int [d\ell]
\prod_{i,a\in I_i} e^{-z_i\ell_a}
\ee
which can be directly evaluated to give
\be
\l w(z_1)\cdots w(z_s) \r\subg = \sum_\Gamma {2^{-2k}\over n(\Gamma)}
\prod_{prop} {2\over z_i+z_j}.
\ee
Here the summation is over all diagrams $\Gamma$ that contribute to $\Mgs$ and
the product is over all propagators. A factor $2/(z_i+z_j)$ is included when
the two sides of the propagator form part of the loops $C_i$ and $C_j$. This is
but a small modification of the Feynman rules of the integral (\ref{count}). We
only have to change the quadratic part. The factor $2^{-2k}$ can be absorbed in
the weight of the cubic vertex, or in the string coupling constant, if one
wishes. In fact, we arrive in this way directly
at a representation of the generating functional of the loop operator
correlation functions
\be
\tau(z_1,\ldots,z_N) = \exp\sum_{g=0}^\infty\l \exp \sum_{i=1}^N w(z_i)\r\subg
\ee
by means of the matrix integral
\be
\tau(Z) = \rho(Z)\inv \int dY \cdot \exp \,Tr[-{1\over2}ZY^2+{1\over6}Y^3].
\label{matrix-int}
\ee
Here $z_1,\ldots,z_N$ are the eigenvalues of the $N \times N$
Hermitian matrix $Z$. We have divided by the one-loop integral
\be
\rho(Z) = \int dY \cdot \exp -{1\over2}
\Tr ZY^2 =\prod_{i,j} (z_i+z_j)^{-{1\over2}}
\ee
in order to obtain the correct asymptotic expansion. We see from (\ref{wz})
that (\ref{matrix-int}) corresponds to the following parametrization of the
coupling constants $t_k$
\be
t_k=\sum_{i=1}^N {1\over k} z_i^{-k}={1\over k} \Tr Z^{-k}.
\ee
We will learn in \S3 that this is a very natural parametrization from the
point of view of the KdV hierarchy. The matrix integral can be made absolutely
convergent by putting a factor of $i$ in front of the cubic term in the action.
One easily sees that, by charge conservation, this is equivalent to a
redefinition $t_k \ra -t_k$. In this way we recover Kontsevich's result
\cite{kontsevich1} that we mentioned in the Introduction.

\newsubsection{Witten's generalizations}

Witten has formulated in \cite{Nmatrix,milnor} a generalization of the above
intersection theory, which involves the choice of an integer $p \geq 2$, and
extra quantum numbers $k_i = 0,\ldots,p-2$ for each of the points $x_i$. Our
observables are now of the form $\sigma_{n,k}$, and their correlation functions
are defined as
\be
\l\sigma_{n_1,k_1} \cdots \sigma_{n_s,k_s} \r\subg \equiv
p^{-g} \Bl \sigma_{n_1}(1) \cdots \sigma_{n_s}(s) \cdot c_D(V),\Mgsbar' \Br
\label{p-cor}
\ee
Let us explain the different objects that enter in this definition. We have
already met the Mumford classes $\sigma_n$. The cohomology class $c_D(V)$ is
the top dimensional Chern class of a complex $D$-dimensional vector bundle $V$
over moduli space. The definition of $V$ depends on the numbers $p$ and $k_i$
as follows \cite{Nmatrix,milnor}. Let $S$ be the line bundle $K^{p-1}
\otimes_i \cO(z_i)^{k_i}$. That is, its section are $p-1$ forms that can have
poles of order $k_i$ at the point $x_i$. The degree of $S$ is given by
\be
\hbox{deg}(S) = (p-1)(2g-2) + \sum_{i=1}^s k_i.
\ee
If this degree is a multiple of $p$ we can define another line bundle $T$ as
the $p\th$ root of $S$
\be
T^{\otimes p} = S,\quad\hbox{if}\quad\hbox{deg$(S)\equiv 0$ (mod $p$).}
\ee
Actually there are $p^{2g}$ choices, and this defines a branched covering of
$\Mgs$ that is denoted as $\Mgs'$. The fiber $V_\Sigma$ of the vector bundle
$V$ at a point $\Sigma \in \Mgs$ is defined as the space of holomorphic
section of $T$
\be
V_\Sigma = H^0(\Sigma,T),\quad\hbox{if}\quad H^1(\Sigma,T)=0.
\ee
This is true if the degree of $T$ is large enough. (See \cite{milnor} for a
more precise definition that takes into account that $H^1(\Sigma,T)$ may not
always vanish.) In that case Riemann-Roch tells us that the fiber of $V$ has
complex dimension
\be
D = d(g-1) + \sum_{i=1}^s q_i,
\label{D}
\ee
where we introduced the new notations
\be
d={p-2\over p},\qquad q_i={k_i\over p}.
\ee
By construction $D$ is always a non-negative integer. Combining all these
ingredients we see that the correlation functions are only non-vanishing if the
following charge conservation holds
\be
\sum_{i=1}^s (n_i+q_i-1)=(3-d)(g-1).
\label{p-charge}
\ee

It is sometimes convenient to denote the operators $\sigma_{n,k}$ slightly
different. From the definition it is clear that $\sigma_{n,k}$ is more or less
the tensor product of a Chern class and `fractional class' related to the
bundle $T$. That is, we would like to write symbolically
\be
\sigma_{n,k} \sim \sigma_n \otimes \phi_k
\label{tensor}
\ee
with $\phi_k$ an operator of fractional charge $q=k/p$. We will make more sense
out of this notation later on, but at this moment we only want to emphasize the
special nature of the fields $\sigma_{0,k}$ that we also write as $\phi_k$.
These operators do not depend on the Mumford classes, only on the top Chern
class $c_D(V)$, and we will refer to these operators as `primary.' The primary
correlation functions
\be
\l\phi_{k_1}\cdots \phi_{k_s} \r\subg \equiv p^{-g} \l c_D(V),\Mgsbar' \r
\ee
can be studied in their one right, and this is actually what we will undertake
in \S6. The primary fields $\phi_k$ can be considered as the observables in a
separate topological field theory, and the combinations with the Mumford
classes can be described as the coupling to topological gravity. This line of
thought suggests many other generalizations, as we will see in the second half
of these notes.

To state the generalized conjectures about the general correlation functions
(\ref{p-cor}) we have again to renormalize our operators, this time as
\be
\cO_{np+k+1} = (np+k+1)((n-1)p+k+1)\cdots(k+1) \cdot \sigma_{n,k}.
\label{calO}
\ee
So, the operators $\cO_n$ in this case are labeled by a positive integer $n$
that satisfies $n \neq 0$ (mod $p$).
The claim is that $\tau(t)$ as defined by
\be
\tau(t)=\exp \sum_{g=0}^\infty \Bl \exp \sum_n t_n \cO_n \Br\subg
\ee
is now a $\tau$-function of the $p\th$ generalized KdV hierarchy. Actually
there is also a simple candidate matrix integral representation of $\tau(t)$
that we will meet in the next section---we basically replace the cubic
interaction by a vertex of order $p+1$. In \S\S6 and 7 we will consider the
quantum field theory that lies behind this conjecture.

It is not difficult to find the generalizations of the puncture and dilaton
equations in this setup. These are again related to the special role of the
operators $\sigma_{0,0}$ and $\sigma_{1,0}$. For instance, one easily obtains
the generalized puncture equation
\be
\l \sigma_{0,0} \cdot \sigma_{n_1,k_1} \cdots \sigma_{n_s,k_s} \r\subg =
\sum_{i=1,\; n_i\neq 0}^s \l \sigma_{n_1,k_1} \cdots \sigma_{n_i-1,k_i} \cdots
\sigma_{n_s,k_s} \r\subg
\label{p-punc}
\ee
The genus zero correction reads in this case
\be
\l \sigma_{0,0} \sigma_{0,i} \sigma_{0,p-i-2} \r\subzero = 1
\label{threepoint}
\ee
or, in the notation (\ref{calO}),
\be
\l \cO_1 \cO_i \cO_j \r\subzero = ij\cdot\delta_{i+j,p}
\label{excep}
\ee
A dilaton equation, completely analogous to (\ref{dilaton}), can also be
derived, with as only new ingredient the expectation value \cite{milnor}
\be
\l \cO_{p+1}\r\subone = (p+1)\l \sigma_{1,0} \r\subone = {p^2-1 \over 24}.
\ee

As a little warm-up for our discussion in \S4.4, we mention here that the
puncture equation (\ref{p-punc}) can be written as a linear partial
differential equation for the string partition function $\tau(t)$
\be
L_{-1} \cdot \tau = 0,
\ee
where $L_{-1}$ is the differential operator
\be
L_{-1} = - \ddt1 + \sum_{k=p+1}^\infty k\,t_k \ddt{k-p} + \half \sum_{i+j=p}
ij\, t_i t_j.
\ee
This identity follows immediately, once it is realized that in a generating
functional a derivative $\d\tau/\d t_k$ represent an insertion of the operator
$\cO_n$ in each correlation function, whereas a multiplication $t_n\cdot\tau$
eliminates the same operator. The last term on the RHS represents the
correction for the exceptional case (\ref{excep}). A little further
experimentation, which we leave as an exercise to the reader, shows that a
particular linear combination of the dilaton equation and the charge
conservation condition (\ref{p-charge}) leads to a similar differential
equation
\be
L_0 \cdot \tau = 0,
\ee
with $L_0$ given by
\be
L_0 = - \ddt{p+1} + \sum_{k=1}^\infty k\,t_k \ddt k + {p^2-1 \over 24}.
\ee
These two equations are but the beginning of a full tower of equations, as we
alluded to in the Introduction, and we will discus this at greater lengths in
\S4.4.

Let us end this brief description with some specializations. The case $p=2$
should reduce to `standard' intersection theory on moduli space, which is
indeed the case, since the bundle $V$ is now zero-dimensional by (\ref{D}).
Another interesting case, although somewhat outside our range of definition, is
the case $p=-1$ \cite{Nmatrix}, where (with all $m_i=0$) $V$ can be identified
with the bundle of quadratic differentials, \ie, the cotangent bundle
$T^*\Mgo$ to moduli space. In this case the top Chern class $c_D(V)$ actually
represents, up to a sign, the Euler class of moduli space and we can calculate
\be
\l 1 \r\subg = \l c_D(V),\Mgobar \r = (-1)^{3g-3}\chi_{g,o},
\ee
where we used the notation $\chi_{g,s}$ for the (virtual) Euler  characteristic
of the moduli space $\Mgs$. Actually due to the work of Harer and Zagier
\cite{harer-zagier} and, in particular, Penner \cite{penner} we know a
concrete realization of the generating functional of these Euler numbers. If
$Y$ is again an $N\times N$ matrix then the following matrix integral generates
all Euler characteristics
\be
\log \int dY \cdot \exp {1\over\lambda} \Tr[\log(1-Y) -Y]
= \sum_{g,s} N^s \lambda^{2g-2+s} \chi_{g,s}.
\ee
The proof of this result is based on the observation that this matrix integral
simply counts all $k$-simplices in the cell decomposition of $\Mgs$ weighted
with a factor $(-1)^k$.

\newsection{Integrable Hierarchies}

In this section we will change our point of view and investigate in greater
detail the integrable hierarchies that have emerged in solvable string theories
and intersection theory. The following is basically a self-contained elementary
exposition of a number of familiar techniques in the theory of integrable
systems, and some readers may wish to skip this part. A good survey of these
matters is given in \cite{segal-wilson}, which we will follow closely. We
gladly acknowledge that a much higher level of sophistication can be found in
the extensive literature on integrable systems. All these techniques will find
their due applications in \S4.

\newsubsection{The generalized KdV and KP hierarchies}

We will first describe the so-called generalized KdV hierarchy in its
scalar Lax formulation. Let $x$ be a real variable and
let $D$ denote the operator
\be
D = \dd x .
\ee
For a given integer $p \geq 2$ we will consider the general
differential operator of order~$p$
\be
L = D^p + \sum_{i=0}^{p-2} u_i(x) D^i.
\ee
Here the coefficients $u_i(x)$ are {\it a priori} arbitrary
functions in the variable $x$.
Note that by conjugation the operator $L$ can
always be put in the above form with vanishing term of order $p-1$. We now
consider the flow of $L$ in an infinite set of `times' $t_1,t_2,\ldots$
as generated by Hamiltonians $H_1,H_2,\ldots$
\be
\Ddt n L = [H_n,L] .
\label{flow}
\ee
In order to be able to consider the operator $L(t)$ as a simultaneous
functions of all parameters $t_n$ the Hamiltonians---which in general are
time-dependent themselves---are required to generate commuting flows, and
therefore satisfy the zero-curvature relation, or Zakharov-Shabat equations,
\be
\Ddt n {H_m} - \Ddt m {H_n} + [H_m,H_n]=0.
\ee
There is a unique basis for such commuting Hamiltonians; they are given by
the non-negative part of the fractional powers of $L$
\be
H_n = (L^{n/p})_+ .
\label{ham}
\ee
Let us explain this notation. We can define fractional powers of $L$ as Laurent
series in the differential $D$. So in general these are so-called
{\it pseudo}-differential operators, \ie , operators of the form
\be
A = \sum_{i=-\infty}^{n} a_i(x) D^i .
\label{laurent}
\ee
The restriction of this sum to only the non-negative powers of $D$ is denoted
by $A_+$
\be
A_+ = \sum_{i=0}^n a_i(x) D^i .
\ee
One similarly defines $A_- = A-A_+$. We take this occasion to introduce another
useful concept: the {\it residue} of a pseudo-differential operator. This is
defined as the coefficient of $D\inv$ in the Laurent expansion (\ref{laurent})
of $A$
\be
\res A = a_{-1}(x).
\ee
We will apply this notion in a moment.

It is not difficult to explicitly verify that the flows (\ref{ham}) indeed
commute. These hierarchies, describing the evolution of an $p\th$ order
differential operator, are known as the generalized KdV or Gelfand-Dikii
\cite{gelfand-dikii} hierarchies. Since the operator $L$ satisfies
(\ref{flow}), its coefficients can be considered to be functions of both the
coordinate $x$ and the times $t_n$. In fact, this is a somewhat redundant
parametrization, because $x$ can be identified with $t_1$, since we have
\be
H_1 = (L^{1/p})_+ = D .
\ee
Furthermore, if $n$ is a multiple of $p$, the flows $t_n$ are trivial
\be
\Ddt n L =0, \qquad \hbox{if $n= q \cdot p$,}
\ee
since in that case $H_n = L^q$ which clearly commutes with $L$.

As an example we can consider the simplest case $p=2$, where we have
\be
L = D^2 + u(x),
\ee
the Hamiltonian of a particle in a potential $u(x)$. One arrives
at the original Korteweg-de Vries equation by considering the first non-trivial
flow of this operator, as generated by $H_3=D^3+{3\over2}uD+{3\over4}u'$,
\be
\Ddt 3 L = \Ddt 3 u = [H_3,L] = u \Dd x u + {1\over 6}{\d^3 u \over \d x^3} .
\ee

All the KdV flows can be considered special cases of the more general
KP hierarchy which is defined as follows. Let $Q$ denote the unique
$p\th$ root of the differential operator $L$
\be
Q = L^{1/p},
\ee
that has an expansion of the form
\be
Q=D + \sum_{i=1}^\infty q_i D^{-i}.
\label{visje}
\ee
Then $Q$ satisfies a similar set of evolution equations as $L$
\be
\Ddt n Q = [H_n,Q],\qquad H_n=Q^n_+.
\ee
These can be considered as the defining equations of the KP hierarchy, if $Q$
is now taken to be an {\it arbitrary} pseudo-differential operator with an
expansion (\ref{visje}). One specializes to the $p\th$ KdV case by imposing
that $Q^p=L$ is a pure differential operator, that is, by requiring
\be
Q^p_- = 0.
\ee
The KP hierarchy can be slightly rewritten by introducing yet another
pseudo-differential operator $K$ through
\be
Q = K D K\inv,
\label{K}
\ee
where $K$ has an expansion of the form
\be
K = 1 + \sum_{i=1}^\infty a_i D^{-i}.
\ee
It satisfies the equation
\be
\Ddt n K = - Q^n_-\cdot K.
\ee

We end this long list of definitions with a side remark. It is rather
straightforward to take the `classical' limit, also known as the {\it
dispersionless} KdV hierarchy \cite{zakharov,krichever} (see also the lectures
of I.~Krichever in these proceedings.) We simply put $\hbar$ in all the
obvious places and take the $\hbar \ra 0$ limit. In this fashion the
differential $D$ becomes the classical momentum $y$ conjugate to $x$, and the
differential operator $L$ becomes a function, polynomial in $y$, on the phase
space $(y,x)$ with symplectic form $dy\wedge dx$,
\be
L(y,x) = y^p + u_{p-2}(x) y^{p-2} + \ldots + u_0(x).
\ee
All commutators reduce to Poisson brackets, in particular we have
\be
\Ddt n L = \{H_n,L\} = \Dd y {H_n} \Dd x L - \Dd x {H_n} \Dd y L.
\ee
We will see that, in terms of the string partition function, $\hbar$
corresponds to the string coupling constant $\lambda$, and that the classical
limit implies the restriction to genus zero. We will return to these matters
in \S6.4.

\newsubsection{The $\tau$-function and the Baker-Akhiezer function}

To a solution $L(t)$ of the KdV equations (\ref{flow})
one can associate a so-called
{\it tau-function} $\tau(t)$. It is in several ways related to the differential
operator $L$. A particularly concrete connection, though perhaps not the most
insightful, is given in terms of the operator $K$ defined in (\ref{K})
\be
\res K= - \dd x \log \tau,
\label{resK}
\ee
or, equivalently, in terms of the Lax operator $L$
\be
\res L^{n/p}= {\d^2 \over \d x \d t_n} \log \tau .
\label{resL}
\ee
Note that if $i=1,\ldots,p-1$,
\be
\res L^{i/p}= {i\over p}\cdot u_{p-i-1} + \ldots
\label{lu}
\ee
where the ellipses indicate polynomials in the coefficients $u_j$ and there
derivatives $u_j',u_j'',\ldots$ with respect to $x=t_1$, all satisfying $j
>p-i-1$. Thus the transformations from the coefficients $u_{i-1}$ to the
residues $\res L^{i/p}$ is seen to be upper-triangular and thus invertible. In
this sense the operator $L$ can be reconstructed from the $\tau$-function.

The opposite problem, how to construct $\tau$ out of $L$, will be considered in
greater detail in \S 3.4. A crucial ingredient in this construction is played
by the eigenfunctions $\psi$ of $L$. Since the operator $L$ depends itself on
the times $t_n$, these eigenfunctions depend on the coordinate $x=t_1$, the
other variables $t_n$ and the eigenvalue that we choose to parametrize as
$z^p$. With the understanding that $x$ can be identified with $t_1$ we will
write the eigenfunction as $\psi(t,z)$. Its defining equation is
\be
L \, \psi(t,z) = z^p \,\psi(t,z).
\label{baker}
\ee
The function $\psi(t,z)$ is called the {\it Baker-Akhiezer function.}
It satisfies the Schr\"odinger equations
\be
\Ddt n \psi = H_n \psi.
\ee
We will assume it is normalized such that
\be
\psi(t,z) = g(t,z) \cdot \eta(t,z),\qquad g(t,z)=\exp \sum_n t_n z^n,
\ee
where $\eta(t,z)$ has an asymptotic expansion in $z\inv$ of the form
\be
\eta(t,z) = 1 + \sum_{i=1}^\infty a_i(t) z^{-i}.
\label{eta}
\ee
Note that $\eta=1$ corresponds to the trivial case $L=D^p$. In fact, since in
the general case $L=KD^pK\inv$, one can write $\psi$ as
\be
\psi(t,z) = K \cdot g(t,z)
\ee
which shows that the coefficients $a_i$ in (\ref{K}) and (\ref{eta}) are indeed
identical.

\newsubsection{Grassmannians and fermions}

There is a beautiful way to generate solutions to the KP hierarchy using
Grassmannians or equivalently two-dimensional free (chiral) fermions. This
method, originally due to Sato \cite{sato,date}, is further explored and
exposed in the beautiful paper by Segal and Wilson \cite{segal-wilson}.

The basic idea is the following. For every operator $L(t)$ we can consider its
Baker-Akhiezer function $\psi(t,z)$ as defined in (\ref{baker}). For fixed
value of the $t_n$'s, and after an appropriate analytic continuation in $z$,
the function $\psi(z)$ can be seen as a wave function in the Hilbert space
$H=L^2(S^1)$ with $z$ the coordinate on the unit circle. When the function
$\psi(t,z)$ evolves in time it will sweep out a linear subspace $W$ of $H$
\be
\psi(t,z) \in W \subset H.
\ee
The method of Grassmannians constructs the solution $\psi(t,z)$ out
of a given linear space $W \subset H$.

The wave functions in $H$ can be given the interpretation of one particle
fermion states, with Hamiltonian $z \d_z$. We have a basis of eigenstates
$\{z^n\}$ in $H$ with energy $n\in \Z$, and a natural polarization
\be
H=H_+\oplus H_-
\ee
in terms of positive and negative energy states. That is, $H_+$ has a basis
$z^0,z^1,\ldots,$ whereas $H_-$ is spanned by $z^{-1},z^{-2},\ldots$. The
Grassmannian $Gr(H)$ is defined as the space of subsets $W$ that (in some
precise sense) are comparable to $H_+$. The Grassmannian can be seen as the
second quantized fermion Fock space. Indeed, to the subspace $H_+$ we can
associate the semi-infinite wedge product
\be
|0\r = z^0 \wedge z^1 \wedge z^2 \wedge \ldots
\ee
which is just the Dirac vacuum, where---with an unfortunate choice
of conventions---we filled all `positive energy states.'
If $w_0,w_1,\ldots$ is a basis for the subspace $W$, we can similarly
construct the state
\be
|W\r = w_0 \wedge w_1 \wedge w_2 \wedge \ldots
\ee
Recall that in this notation the fermion fields have an expansion
\be
\psi(z) = \sum_n \psi_n z^{n},\qquad
\psi^*(z) = \sum_n \psi^*_n z^{n},
\ee
and act on the states as
\be
\psi_n = \dd {z^n},\qquad \psi^*_n = z^{-n} \wedge.
\ee
These operators obviously satisfy the canonical anti-commutation relations
\be
\{\psi_n,\psi_m\}=\{\psi^*_n,\psi^*_m\}=0,
\qquad\{\psi_n,\psi^*_m\}=\delta_{n+m,0}.
\ee
We want the Grassmannian $Gr(H)$ to consist of
subspaces $W$ such that the inner product of the states corresponding
to $H_+$ and $W$ is well-defined. Recall that
\be
\l 0 | W \r = \det \l w_i,z^j\r
\ee
in terms of the one-particle inner product $\l\cdot,\cdot\r$, so $W$ and $H_+$
must be comparable in size. This will certainly be the case if $W$ is {\it
transverse} to $H_-$. By this we mean the following. Let $w$ be the map $H_+
\ra W$ that sends the basis elements $z^i$ of $H_+$ to the basis elements
$w_i$ of $W$
\be
z^i \ra w_i = \sum_j (w_+)_{ij}z^j + \sum_k (w_-)_{ik} z^{-k}.
\ee
This defines for us the projections $w_+$ and $w_-$. We now have
\be
\l 0 | W \r = \det w_+
\ee
We will require that $w_+$ is invertible. In this case
$W$ can be seen as the graph
of the map $A= w_+^{-1}w_-$, and we can choose a basis
\be
w_i = z^i + \sum_{j=1}^\infty A_{ij} z^{-j}.
\label{basis}
\ee
Clearly the case $A=0$ corresponds to $W=H_+$. In terms of matrix $A_{ij}$ the
state $|W\r$ can be written as the Bogoliubov transform
\be
|W\r = h |0\r,\qquad h=\exp \sum_{i,j} A_{ij}\psi_{i}\psi^*_{j},
\ee
with $h$ an element of a subgroup of $GL(\infty,\C)$ that has a natural action
on the Grassmannian $Gr(H)$.

There is a smaller subgroup $\Gamma_+ \subset GL(\infty,\C)$ defined by
multiplication with a function $g(z)$ that is holomorphic on the unit disk
$D_0$. A map $g(z)\in\Gamma_+$ simply multiplies every element $w \in W$. With
respect to the polarization $H_+ \oplus H_-$ of $H$ it has a decomposition of
the form
\be
g = \twomatrixd a b 0 c
\label{A}
\ee
We can now consider the orbit $\Gamma_+ \cdot W$ of a subspace $W \in Gr(H)$
under the action of $\Gamma_+$. If we parametrize $g(z)$ as
\be
g(t,z) = \exp \sum_{n=1}^\infty t_n z^n \label{gt}
\ee
the evolution operator $U(t)$ that acts in the fermion Fock space is defined by
\be
U(t) |W\r = |gW\r.
\ee
Multiplication by $z^n$ simply shifts $z^k \ra z^{k+n}$, so $U(t)$ is realized
in terms of the fermion fields as
\be
U(t) = \exp \sum_{n=1}^\infty t_n J_n .
\ee
Here $J_n$ are the modes of the fermion current $J(z) = \psi\psi^*(z)$, with
the fundamental property $[J_n,\psi_k]=\psi_{k+n}$. We can write $U(t)$ as
\be
U(t) = \exp \oint {dz\over 2\pi i}\log g(z) \cdot J(z).
\label{U}
\ee

We are now finally in a position to explain how all this is related to the Lax
pair formulation of the integrable hierarchies that we considered in the
previous subsections. The subspace $W$ has, by definition, a unique element
$\eta = w_0$ satisfying
\be
\eta(z) = 1 + \sum_{i=1}^\infty a_i z^{-i}.
\ee
It can be described formally as the intersection
\be
\eta(z) = W \cap (1+H_-).
\ee
Similarly the transform $g\inv W$ contains an element with that property
\be
\eta(t,z) = g\inv W \cap (1+H_-).
\ee
Recall here that $g$ is parametrized as in (\ref{gt}). But this last equation
implies directly that
\be
\psi(t,z) = g(t,z)\cdot \eta(t,z) \in W.
\ee
Since this is true for {\it all} times $t_n$ and $W$ is a linear space, we have
the property that all (multiple) derivatives of $\psi$ still lie in $W$. In
particular we have
\be
g\inv \Ddt n \psi = z^n + \ldots \in g\inv W,
\ee
and
\be
g\inv {\d^k \psi \over \d x^k} = z^k + \ldots \in g\inv W.
\ee
Since $g\inv W$ has a unique basis of the form (\ref{basis}), this implies
that $\psi$ satisfies an equation
\be
\Ddt n \psi = H_n \psi.
\ee
where $H_n$ is an $n\th$
order differential operator in $x$. By construction the $H_n$ commute, and thus
we have found a solution to the KP hierarchy. The reduction to KdV comes about
when $W$ satisfies the extra constraint
\be
z^p \cdot W \subset W.
\ee
In that case one finds the additional relation \be L \psi = z^p \psi, \ee which
defines the differential operator $L=H_p$.

\newsubsection{$\tau$-functions and determinants}

The tau-function $\tau(t)$ associated to a point $W$ in the Grassmannian is
now defined as
\be
\tau(t)= {\det (g\inv w)_+ \over g\inv \det w_+}
\ee
with $g$ parametrized as in (\ref{gt}), or in our notation
(\ref{basis})-(\ref{A})
\be
\tau(t)= \det (1+ a\inv b A).
\ee
In terms of the fermion Fock space we have the representation
\be
\tau(t)={\l 0| U(t) |W \r \over g\inv \l 0|W \r}
\ee
Since the state
\be
\l t|=\l0|U(t)
\ee
can be considered as a coherent state in the Hilbert space, the $\tau$-function
can alternatively be regarded as the wave-function of the state $|W\r$.

The bosonization formulas
\be
J= \d \phi,\quad \psi=e^\phi,\quad \psi^*=e^{-\phi},
\ee
lead us to consider the operators
\be
\gamma_+(z) =\exp \sum_{n=1}^\infty {1\over n} z^{-n}\ddt n .
\ee
Insertion of the fermi field $\psi(z)$ corresponds to action on $\tau(t)$
with the differential operator
\be
\gamma(z) \equiv \gamma_+(z) \cdot g(z),
\ee
together with an increase in fermi number of one unit.
If we denote the $N$-particle ground state as
\be
|N\r = z^{N}\wedge z^{N+1}\wedge z^{N+2} \wedge \ldots
\ee
then we evidently have
\be
\gamma(z_1) \cdots \gamma(z_N) \cdot \tau(0) =
\l N | \psi(z_1) \cdots \psi(z_N) | W \r
\label{N-point}
\ee
where the notation on the LHS
indicates that we first take derivatives and then put the
arguments $t_n=0$ in the $\tau$-function. This implies in particular that
\be
\gamma_+(z_1)\cdots \gamma_+(z_N) \cdot \tau(0) =
\Delta(z)\inv \l N | \psi(z_1) \cdots \psi(z_N)| W \r
\label{bloop}
\ee
with
\be
\Delta(z) = \l N | \psi(z_1) \cdots \psi(z_N)| 0 \r =
\prod_{i>j}(z_i-z_j) = \det\, z_i^{j-1},
\ee
the Vandermonde determinant. In fact, (\ref{bloop}) can be easily evaluated,
since only the first $N$ one-particle states $w_1,\ldots,w_N$ contribute to
give
\be
\l N | \psi(z_1) \cdots \psi(z_N)| W \r
= \det (w_{j-1}(z_i)) \equiv \Delta(w;z).
\ee
Here we introduced the `generalized Vandermonde' determinant
$\Delta (w;z)$ for an arbitrary basis $\{w_k(z)\}$. Note that in this language
the Baker-Akhiezer function is recovered as the fermion one-point function
\be
\psi(t,z) = \gamma(z) \log \tau(t)
\ee
By expanding $\psi$ in powers of $z\inv$ we find as a corollary that our
definition of $\tau$ coincides with the one given in \S3.2.

These fermionic correlation functions occur naturally if we consider special
elements $g(z) \in \Gamma_+$ such that $g\inv$ is a $N\th$ order polynomial
with zeroes $z_i$ outside $D_0$
\be
g(z)\inv = \prod_{i=1}^N (1-z/z_i).
\ee
We can regard $g\inv$ to be the characteristic polynomial
\be
g(z)\inv = \det(1-z \cdot Z\inv),
\ee
where $Z$ is an Hermitian matrix with eigenvalues $z_1,\ldots,z_N$.
This corresponds to the choice of parameters $t_n$ as
\be
t_n = {1\over n} \sum_{i=1}^N z_i^{-n} = {1\over n} \,\Tr Z^{-n},
\label{miwa}
\ee
also known as `Miwa's coordinates' \cite{miwa}.
We will write the tau-function in this parametrization as
\be
\tau(t) = \tau(Z).
\ee
For this particular choice of $g$ we find immediately with (\ref{U}) that
\be
\tau(Z) = \gamma_+(z_1)\cdots \gamma_+(z_N) \cdot \tau(0),
\ee
which we already evaluated to be ratio
\be
\tau(Z) = \Delta(w;z)/\Delta(z).
\ee
This is our main results of this section. It gives a closed expression for the
$\tau$-function in the specific parametrization (\ref{miwa}) once we know an
explicit basis $w_k(z)$ for the point $W \in Gr(H)$.

\newsection{Matrix Integrals}

We will now turn to the application of the results obtained in the previous
section to the specific solutions of the KdV hierarchies that appear in
theories of quantum gravity. In this section we will perform the `Wick
rotation' $x \ra i x$, $t_n \ra i t_n$, in order to achieve better convergence
properties of many of our quantities. This change is of course completely
irrelevant on the level of asymptotic expansions. Note that the derivative
$D=-i\d/\d x$ is now an Hermitian operator. We will assume that the Lax
operator
$L$ has (possibly complex) coefficients such that it is Hermitian too. It
consequently has real eigenvalues $z^p$.

\newsubsection{The string equation and Airy functions}

We can think of the many different solutions to the KdV hierarchies, such as
the famous soliton solutions, as being related to different initial conditions
$L(0)$ for the Lax operator $L(t)$. For the solutions that appeared in \S2
these initial conditions can be found as follows. Recall that in these cases
the $\tau$-function has a very concrete interpretation: it is given by the
string partition function of the $(p,1)$ minimal topological model. Stated
otherwise, the logarithm of $\tau$ is a generating functional for all
correlation functions on connected surfaces of arbitrary genus~$g$, and
consequently has an expansion
\be
\log \tau = \sum_{g=0}^\infty \;\Bl \exp \sum_n t_n {\cal O}_n \Br\subg
\ee
If we think of the variables $t_n$ as describing a background, then $\log \tau$
becomes indeed the `string free energy'
\be
\log \tau = F = \l 1 \r\subt
\ee
where the subscript $\l\cdot\r_t$ simply indicates the insertion of the
`$\exp t_n\cO_n$' term and a summation over all genera is understood.
In this notation equations (\ref{resK})--(\ref{resL}) relate the residues with
special two-point functions
\be
\res K = - \l \cO_1 \r\subt \qquad
\res L^{n/p} = \l \cO_1 \cO_n \r\subt \label{res-cor}
\ee
We now want to determine the initial condition $L(0)$, where we have put all
times $t_n$ to zero for $n>1$. Putting the coupling constant $t_1=x$ to a
non-zero value and expanding $\log\tau$ up to first order in the other coupling
constants, we find for $i=1,\ldots,p-1$ with (\ref{threepoint})
\be
\res L^{i/p} = p\cdot\delta_{i,p-1}.
\ee
(The factor $p$ that appears here can be absorbed in the string coupling
constant, and we will drop it in the consequent.)
This translates in the following initial value for the differential operator
$L$
\be
L(0) = D^p + x.
\ee
That is, only a linear potential $u_0(x)=x$ and no higher derivative terms
appear. The equation
\be
u_i(x) = x \cdot \delta_{i,0},
\ee
or, equivalently,
\be
[D,L]=1,
\ee
is known as the {\it string equation} for the $(p,1)$ model, in the notation of
the introduction.\footnote{For general $(p,q)$ it reads $[H_q,L]=1$
\cite{douglas}, which in the case $(p,q)=(2,3)$ gives the celebrated Painlev\'e
equation $u^2+{1\over6}u''=x$ of \cite{3}.} It is this extremely trivial form
of the string equation that leads to the complete solvability. Indeed, we will
now be able to explicitly solve for the $\tau$-function, because in this case
the Baker-Akhiezer function
\be
L \, \psi(x,z) = z^p \,\psi(x,z)
\ee
is simply given by the generalized Airy function
\be
\psi(x,z) = c(z) \cdot
\int dy \, \exp i[(x-z^p)y+{y^{p+1}\over p+1}].
\label{airy}
\ee
This should be contrasted with the much more complicates solutions for an
arbitrary $(p,q)$ model \cite{moore}. The coefficient $c(z)$ is fixed by
requiring the asymptotic expansion (\ref{eta})
\be
\psi(x,z) = e^{ixz}\left(1 + \sum_{i=1}^\infty a_i(x) z^{-i}\right).
\label{jeroen}
\ee
The normalization can be chosen as
\be
c(z) = e^{{ip\over p+1}z^{p+1}} \sqrt{z^{p-1}}.
\label{c}
\ee
The asymptotic expansion is now verified by shifting $y \ra y+z$ in the
integral (\ref{airy}), so that one obtains the representation
\be
\psi(x,z) = e^{ixz} \sqrt{z^{p-1}} \int dy \cdot e^{ixy+iS(y,z)}
\ee
with
\be
S(y,z)= {1\over p+1} \left[(y+z)^{p+1}\right]_{\geq 2}.
\label{S}
\ee
Here the subscript `$\geq2$' implies that we only keep terms in $y$ of second
order or higher. By using the stationary phase approximation to such integrals
one then easily deduces the expansion (\ref{jeroen}).

We now wish to determine the function $\tau(Z)$ for this special initial value
condition
\be
L=D^p+x.
\label{L}
\ee
According to the general discussion of section 3, the Baker-Akhiezer function
$\psi(t,z)$ is an element of $W$ for all values of $t$. This is in particular
true if we restrict to the parameter $t_1=x$. So, all the functions $v_k(z)$
that appear in the Taylor expansion
\be
\psi(x,z) = \sum_{k=0}^\infty v_k(z) {x^k\over k!}.
\ee
are elements of $W$:
\be
v_k(z) \in W.
\ee
Furthermore, the functions $v_k(z)$ form a basis, since they are given by
\be
v_k(z) = \sqrt{z^{p-1}} \int dy \cdot (y+z)^k \cdot e^{iS(y,z)},
\ee
and thus have to property
\be
v_k(z) = z^k(1 + O(z\inv)).
\ee
They therefore define an element of the Fock space by
\be
|W\r = v_0 \wedge v_1 \wedge v_2 \wedge \ldots
\ee
and the $\tau$-function is, according to the main result of \S3,
simply obtained as the following ratio of determinants
\be
\tau(Z) = \Delta(v;z)/\Delta(z).
\ee

\newsubsection{More general models: the small phase space}

One can slightly generalize the previous case by considering an
operator of the form
\be
L = W(D) + x,
\label{W}
\ee
with $W$ an arbitrary polynomial of order $p$ with {\it constant} coefficients
\be
W(y)= y^p + g_{p-2}y^{p-2} + \ldots + g_0.
\label{potential}
\ee
If we introduce a potential $V$ by the relation
\be
W(y)=V'(y),
\ee
and a function $\zeta(z)$ through
\be
V'(\zeta)=z^p,\quad \zeta=z + O(z\inv),
\ee
the Baker-Akhiezer function reads in this case
\be
\psi(x,z) = e^{ix\zeta} \sqrt{V''(\zeta)} \int dy \cdot e^{ixy+iS(y,\zeta)}
\ee
with $S$ defined as
\be
S(y,\zeta)=\left[V(y+\zeta)\right]_{\strut \geq2} =
V(y+\zeta)-V(\zeta)-y V'(\zeta).
\ee
In this case the moments $v_k(z)$ are given by
\be
v_k(z) = \sqrt{V''(\zeta)} \int dy \cdot (y+\zeta)^k \cdot e^{iS(y,\zeta)}
\ee
This case is actually of great interest, since an operator of the form
(\ref{W}) is naturally obtained if we let the operator $L=D^p+x$ evolve in the
`primary' times $t_1,\ldots,t_{p-1}$. The subspace parametrized by these
$t_i$'s is known as the `small phase space' \cite{dijkwit}. In intersection
theory the restriction to the small phase space implies that we only consider
the top Chern class of the vector bundle $V$, not the Mumford-Morita-Miller
classes $\sigma_n$. In the corresponding quantum field theory the primary
operators $\cO_1,\ldots,\cO_{p-1}$ are the physical fields of a $(p,1)$
topological minimal model. The small phase space describes the moduli space
of topological field theories that can be reached by perturbation of the
minimal models. When we consider $\tau(t)$ as a function on the
small phase space, it receives furthermore, by charge conservation, only
contributions at genus zero.

In the context of integrable hierarchies of KdV type, these special properties
of the first $p-1$ flows are reflected in the fact that the corresponding
Hamiltonians $H_i=L^{1/p}_+$ do not depend on $x$. Therefore the
dispersionless, or spherical, approximation is exact, and we have the
classical equations
\be
\Ddt i L = \{H_i,L\} = \{L^{i/p}_+,x\} = \Dd y {L^{i/p}_+}
\label{top-kdv}
\ee
with $y$ the classical momentum corresponding to $D$. These equations fully
determine the $t$-dependence of $L$. The initial value
\be
W(y)=y^p
\ee
evolves to a general potential of type (\ref{potential}), where the
coefficients $g_i(t)$ are given implicitly by the relation
\be
i(p-i)t_{p-i} = \res L^{i/p} =
\oint {dy\over 2\pi i} \; L(y)^{i/p},\qquad (i=1,\ldots,p-1).
\ee
This equation follows immediately from the results (\ref{res-cor}) and
(\ref{threepoint}).

\newsubsection{Matrix integrals}

We will now show following Kontsevich \cite{kontsevich3}
how the previous results are related to
matrix integrals. First recall the wonderful result of Harish-Chandra
\cite{harish-chandra}, Mehta \cite{metha}, Itzykson and Zuber
\cite{itzykson-zuber}, for the following integral over the unitary group $U(N)$
\be
\int dU \, \exp i\,\Tr [UXU^\dagger Y] = c \cdot {\det e^{ix_iy_j}
\over \Delta(x) \Delta(y)},
\label{hc}
\ee
with $dU$ the Haar measure and $x_i,y_i$ the eigenvalues of the Hermitian
matrices $X$ and $Y$. This result can be obtained in many different ways,
perhaps most elegantly as an application of the Duistermaat-Heckman
localization formula \cite{picken}.

As a special application of the above equation consider the matrix Fourier
transform, \ie , the integral over a hermitian matrix $Y$ in a external field
$X$, both $N \times N$ matrices, of the form
\be
\tau(X) = \int dY\,\exp i\,\Tr [XY + V(Y)].
\ee
By conjugation invariance, this is only a function of the eigenvalues
$x_1,\ldots,x_N$. We can use (\ref{hc}) to integrate
out the angular variables $U$ in the decomposition
\be
Y = U \cdot {\rm diag}(y_1,\ldots,y_N) \cdot U^\dagger,
\ee
which also introduces a Jacobian
\be
dY = \Delta(y)^2 \cdot dU \cdot [dy].
\ee
This leaves us with an integral over the eigenvalues $y_i$ of the form
\be
\tau(X) =
\int [dy] \, \Delta(y)\Delta(x)\inv \exp \sum_j i[x_jy_j + V(y_j)]
\ee
Now the Vandermonde determinant $\Delta(y)$ is a sum of terms of the form
\be
\pm y_1^{i_1}\cdots y_N^{i_N},
\label{yyy}
\ee
and for each of these terms the integral $\tau(X)$ factorizes in separate
integrals over the individual eigenvalues $y_i$. If we introduce the function
\be
w(x) = \int dy \cdot e^{ixy + iV(y)}
\ee
and its derivatives
\be
w_k(x) = \int dy \cdot y^k \cdot e^{ixy + iV(y)},
\ee
the contribution of (\ref{yyy}) is simply
\be
\pm w_{i_1}(x_1)\cdots w_{i_N}(x_N).
\ee
So we can evaluate $\tau(X)$ straightforwardly as
\be
\tau(X) =\det (w_{j-1}(x_i))/ \Delta(x) = \Delta(w;x)/ \Delta(x)
\ee
in the notation of the previous section. This very much suggests that
$\tau(X)$ is a $\tau$-function for the KP hierarchy with one-particle
wave-functions $w_0,w_1,\ldots$. This will be the case if these functions have
appropriate asymptotic expansions, and so depends on the choice of potential
$V$ and the normalization of the integral.

We first wish to apply this result using the functions $v_k(z)$ that appeared
in \S4.1. Recall that these where the derivatives (at $x=0$) of the
Baker-Akhiezer function of the operator $L=D^p+x$. In this case one simply puts
\be
V(Y)={1\over p+1}Y^{p+1},\qquad X=-Z^p,
\ee
 and
\be
\tau(Z) = c(Z)\cdot \int dY\,\exp i\,\Tr [-Z^p \cdot Y + {Y^{p+1}\over p+1}]
\label{int}
\ee
with a proper choice of normalization $c(Z)$. This constant is basically the
product of the individual constants $c(z_i)$ of (\ref{c}), together with a
correction replacing the Vandermonde determinant $\Delta(z^p)$ by $\Delta(z)$
\be
c(Z) = \det Z^{p-1\over 2} \cdot e^{{ip\over p+1}\Tr Z^{p+1}}
{\Delta(z^p) \over \Delta(z)}.
\ee
With all these ingredients the integral (\ref{int})
reproduces indeed our $\tau$-function
\be
\tau(Z)=\Delta(v;z)/\Delta(z)
\ee
The integral can be written in yet another fashion.
Let $S(Y,Z)$ be given as in (\ref{S})
\be
S(Y,Z) = {1\over p+1} \Tr[(Y+Z)^{p+1}]_{\geq2},
\ee
and let $S_2(Y,Z)$ denote the part of the function $S(Y,Z)$ of second order
in $Y$
\be
S_2(Y,Z) = \sum_{k=0}^{p-1}\; \Tr [Y Z^k Y Z^{p-1-k}],
\label{quadr}
\ee
One easily evaluates the Gaussian integral
\be
\rho(Z) \equiv \int dY \, e^{i S_2(Y,Z)} =
\prod_{i,j}\sqrt{ z_i-z_j \over z_i^p-z_j^p}
= \det Z^{1-p\over2} {\Delta(z) \over \Delta(z^p)}.
\ee
This can be used to rewrite $\tau(Z)$ as
\be
\tau(Z) = \rho(Z)\inv \int dY \;e^{i S(Y,Z)}.
\label{tauZ}
\ee
This matrix integral has been suggested as the relevant one for the $(p,1)$
model by several authors, see \eg \cite{kontsevich3,adler,morozov}. Here we
have shown that it follows naturally from the Lax pair formulation of the
string equation, by which the general $(p,q)$ model is characterized.

The integral (\ref{tauZ}) has an obvious asymptotic expansion in matrix Feynman
diagrams. In this way we recover the cell decomposition used by Kontsevich to
derive his results for $p=2$. This elegant geometrical interpretation is less
clear for higher values of $p$. First of all one must include higher order
vertices (up to order $p+1$), which implies the consideration of cells in
$\Mgs$ of non-zero codimension. Secondly, the Feynman rules associated to these
vertices become $Z$-dependent, and generally quite complicated. It would be
very interesting if this prescription could nevertheless be directly related to
the intersection formulas of \S2.4.

As is also observed in \cite{kontsevich3,morozov}, the formula (\ref{tauZ})
makes sense for a much bigger class of potentials. We can consider the general
case
\be
\tau(Z) = c(Z)\cdot \int dY\,\exp i\,\Tr [ -W(Z)\cdot Y + V(Y)]
\ee
If $V(Y)$ has a well-defined Taylor expansion around $Y=Z$, we can put
\be
W(Y)=V'(Y).
\ee
After a shift in the integration variable $Y$, the matrix integral $\tau(Z)$
can be written in the form (\ref{tauZ}), if we define the action to be
\be
S(Y,Z) = \Tr[V(Y+Z)-V(Z)-Y \cdot V'(Z)],
\ee
and $S_2(Y,Z)$ similarly as in (\ref{quadr}) as the quadratic part of $S$.
One easily calculates
\be
\rho(Z) = (\det V''(Z))^{-1/2} {\Delta(z)\over \Delta(V'(z))}
\ee
and the matrix integral (\ref{tauZ}) becomes again a ration of
determinants
\be
\tau(Z)=\Delta(v;z)/\Delta(z).
\ee
In this more general case the functions $v_k(z)$ are given by
\be
v_k(z) = \sqrt{V''(z)} \int dy \cdot (y+z)^k \cdot e^{iS(y,z)},
\ee
and, again under some mild restriction on the potential $V$, this produces a
solution of the KP hierarchy.

One might wonder if all these solutions are completely independent. This brings
us to the following point. The most general matrix integral that naturally
leads to a $\tau$-function of the KP hierarchy depends both on the choice of
potential $V$ and the external field $Z$. This might be a redundant
parametrization, in the sense that a particular variation $\delta V$
corresponds to a flow in the times $t_n$ encoded by $Z$. We meet a simple
example of this phenomenon if $V(Y)$ is a polynomial of order $p+1$. In this
case the functions $v_k$ that we constructed above are identical to the ones
obtained in \S4.2, after we perform a reparametrization
\be
z \ra \zeta(z),\qquad \zeta^p=W(z).
\label{repara}
\ee
Recall that we considered at that point the $\tau$-function associated to the
Lax operator $L=W(D)+x$. A Lax operator of that form could be obtained, if we
start with $W(Y)=Y^p$ and flow in the primary times $t_1,\ldots,t_{p-1}$.
Indeed, as we showed, in that case one ends up with a general polynomial
$W(Y)$ of order $p$ whose coefficients are determined through the relation
\be
t_{p-i} = {1\over i(p-i)} \res W^{i/p},\qquad i=1,\ldots,p-1.
\label{extra}
\ee
We now see that the correct identification of the KP times $t_n$ is given by
\be
t_n = {1\over n} \Tr W(Z)^{-n/p},
\label{redef}
\ee
which just expresses the reparametrization (\ref{repara}) of the coordinate
$z$. Summarizing, the change in potential can be absorbed in the redefinition
in the coupling coefficients (\ref{redef}), together with the extra term
(\ref{extra}) for the primary couplings. Of course, it is possible to consider
more general flows, where the order of the potential changes. This cannot be
absorbed in a shift of the KdV times, and in this way one can interpolate
between different values of~$p$.

As a final application, we can consider the limit $p\ra -1$. In that case we
find
\be
W(Y)=Y\inv,\qquad V(Y)=\log Y.
\ee
One easily verifies that the Kontsevich matrix model now becomes identical to
Penner's model that computes the Euler characteristic of moduli space, in
accordance with our considerations of \S2.4

\newsubsection{The Virasoro constraints}

We will make here a few brief comments on the occurrence of Virasoro algebras
(and more general $W$-algebras) in the context of these integrable hierarchies.
We already mentioned this characterization of the tau-function in the
introduction and we will see in \S7 that it occurs naturally in the context of
topological string theory. There exists by now a quite extensive literature on
this subject, see \eg our selection \cite{virasoro}, and we will be very brief
at this point.

Let us start with a few definitions. In the free fermion theory there is a
holomorphic stress tensor
\be
T(z) = \half(\d_z \psi^* \psi - \psi^*\d_z\psi) = \half J^2,
\ee
which describes the behaviour of the quantum theory under reparametrizations $z
\ra w(z)$ of the unit circle. Its modes $L_n$, defined by
\be
T(z) = \sum_n L_n z^{-n-2},
\ee
satisfy the $c=1$ Virasoro algebra
\be
[L_n,L_m]=(n-m)L_{n+m} + {1\over12}n(n^2-1)\delta_{n+m,0},
\ee
and are the infinitesimal generators of Diff$(S^1)$.
They can be represented as differential operators acting on the $\tau$-function
through the relation
\be
L_{n_1}\cdots L_{n_k} \cdot \tau(t) = \l t|L_{n_1}\cdots L_{n_k}|W\r.
\ee
This gives, with $n>0$, the explicit realizations
\ba
L_{-n} \iss \sum_{k=n+1}^\infty k\,t_k \ddt {k-n} + \half
\!\sum_{i+j=n} \! ij \,t_it_j,\nonu
L_0 \iss \sum_{k=1}^\infty k\,t_k \ddt k,\nonu
L_n \iss \sum_{k=1}^\infty k\,t_k \ddt {k+n} +
\half \!\sum_{i+j=n} {\d^2 \over \d t_i \d t_j}.
\ea
In a $\tau$-function of the $p\th$ KdV hierarchy all $t_k$'s with $k\equiv 0$
(mod $p$) can consistently be put to zero, and only the modes of the form
$L_{q\cdot p}$ act non-trivially.

In the previous sections we have constructed a special solution of the $p\th$
KdV hierarchy, that turns out to have remarkably simple properties with respect
to the Virasoro generators. The results can be formulated as follows. Let us
introduce the redefined operators $L'_n$ by
\be
p \cdot L'_n = L_{n\cdot p}-{\d\over \d t_{1+(n+1)p}} + \delta_{n,0}
{p^2-1\over 24},\qquad n \geq -1.
\ee
One easily verifies that they satisfy the algebra
\be
[L'_n,L'_m]=(n-m)L'_{n+m}.
\ee
The $\tau$-function of the $(p,1)$ model can now be shown to obey the equations
\be
L'_n \cdot \tau =0,\qquad\quad n \geq -1.
\ee
A derivation of this result is given in \cite{loop}. These equations have a
straightforward interpretation as recursion relations for correlation
functions, as we will see explicitly for the case $p=2$ in \S7.

There exists actually a much bigger algebra of currents bilinear in the
fermions---the so-called $W_{1+\infty}$ algebra \cite{w-infty}, generated by
currents of the form
\be
W(z) = \d_z^i\psi^* \d_z^j\psi.
\ee
The modes of this fields can also be realized as differential operators,
and can be considered as the infinitesimal generators of the
group $GL(\infty,\C)$ that acts naturally on Sato's Grassmannian.
The $\tau$-function of the $(p,1)$ model can be shown to satisfy the more
general constraints
\be
W^{(s)}_n \cdot \tau= 0,\qquad s=2,\ldots,p,\ n \geq 1-s,
\ee
where $W_n^{(s)}$ are the modes of a spin $s$ generator $W^{(s)}(z)$. See \eg
\cite{jacob,w-infty} for more details. The fields $W^{(s)}$ generate a
so-called $W$-algebra associated to the Lie group $SL(p)$.

These constraints can actually be found directly in the matrix model
representation of the $\tau$-function \cite{witten-kontsevich}. A very elegant
way \cite{kontsevich2,newman,russians} proceeds as follows. Recall that for the
$(p,1)$ models the $\tau$-function was, up to normalization, given by the
generalized matrix Airy function
\be
A(X) = \int dY \cdot \exp i\,\Tr[XY + {Y^{p+1}\over p+1}].
\ee
This function satisfies the `matrix Airy equation'
\be
\left({\d^p \over \d X^p}+ X\right)A=0,
\ee
a reflection in matrix terms of the `string equation' $L=D^p+x$. When this
equation is written out in terms of the coordinates $t_n$, its gives rise to a
set of equations of the form
\be
Q \cdot \tau = 0.
\ee
where $Q$ is a differential operators of order $p$ or less. The claim is, that
these equations reproduces exactly the Virasoro and $W$-constraints.

\newsection{Topological Field Theory}

We now shift our perspective completely and start our discussion of topological
field theories \cite{tft,sigma}. The models we consider in this and the next
section can be considered as `matter theories' and only describe the primary
fields $\cO_1,\ldots,\cO_{p-1}$ that featured in our previous models.
Consequently, we will only recover the KdV equations on the small phase space.
In order to extend our discussion to the full integrable hierarchies these
models have to be coupled to (topological) gravity---a subject that we will
postpone until \S7, and then only discuss marginally.

\newsubsection{The stress-energy tensor}

Let us start with some remarks of a rather general nature. For other
introductions see \eg \cite{witten-trieste}-\cite{review}. Given a compact
manifold $M$ of dimension $D$ and a quantum field theory with some set of
fundamental fields $\phi(x)$, one can consider the vacuum amplitude or
partition function
\be
Z(M) = \int [d\phi]\; e^{-S[\phi]}
\ee
Here we formally defined $Z(M)$ by the path-integral over all field
configurations on $M$ weighted with some action $S$. In general the partition
function $Z(M)$ will depend on many geometrical data. For instance, in almost
all quantum field theories we need a Riemannian structure on $M$, \ie, a
metric $g_{\mu\nu}(x)$. This metric enters both in the definition of the action
$S$ and in the definition of the measure $[d\phi]$. Other possible ingredients
might be an orientation and, if fermions are involved, a choice of spin
structure, or in the case of gauge fields a choice of fiber bundle.

Although through its definition the quantum field theory can seem to depend
on all information encoded in the metric $g_{\mu\nu}$,
it might be the case that the actual amplitudes
have some invariance under particular changes $\delta g_{\mu\nu}$ in the
metric. An example is the invariance under reparametrizations
\be
\delta g_{\mu\nu} = D_\mu \epsilon_\nu + D_\nu \epsilon_\mu.
\ee
As is well-know, this leads to the conservation law
\be
D^\mu T_{\mu\nu}= 0
\ee
for the expectation value of the stress-energy tensor
\be
T_{\mu \nu} = {1\over \sqrt g}{\delta S \over \delta g^{\mu \nu}},
\ee
that encodes the reaction of the quantum field theory to metric
deformations. Another example of a symmetry is conformal invariance:
the invariance under local rescalings of the metric
\be
\delta g_{\mu\nu} = \epsilon \cdot g_{\mu\nu},
\ee
with $\epsilon(x)$ an arbitrary function.
This implies that the trace of $T_{\mu\nu}$ vanishes
\be
T^{\mu}{}_{\mu}=0.
\ee
Quantum field theories with this property are known as conformal field
theories.

A quantum field theory will be called {\it topological}, when it is invariant
under arbitrary smooth deformation of the metric:
\be
\delta g_{\mu\nu} = \epsilon_{\mu\nu}.
\ee
In this fashion only the topology of the manifold $M$ will matter,
and the partition
function $Z(M)$ will be a topological invariant.
In this case the stress tensor will have to vanish completely
\be
T_{\mu\nu} =0.
\ee
This discussion
can be generalized from vacuum amplitudes to physical correlation functions.
In a generic quantum field theory, the correlation functions depend on the
position of the operators. But in a topological field theory the physical
correlation functions of local operators are just numbers,
\be
\l \phi_{i_1}(x_1)\cdots \phi_{i_s}(x_s) \r\subM
\equiv \int [d\phi]\; \phi_{i_1}(x_1)\cdots \phi_{i_s}(x_s) \cdot
e^{-S}= constant.
\label{correl}
\ee
This is again a manifestation of the metric-independence of physical
observables, and a consequence of the vanishing of the stress-energy tensor,
now within correlation functions
\be
\l T_{\mu \nu}(x) \phi_{i_1}(x_1) \cdots \phi_{i_s}(x_s)\r\subM = 0
\label{e-mvanish}
\ee
In \S5.4 we will see how these conditions can be realized in
practice.

\newsubsection{Factorization and Atiyah's axioms}

We will at this point not be interested in actually evaluating the
path-integral for a given quantum field theory. We will rather, following
Atiyah \cite{atiyah}, deduce from the above definition certain properties of
partition functions, and more generally correlation functions, that can be
lifted to the status of `axioms.' The most important property is the concept of
factorization.

Let us in our imagination cut the manifold $M$ along a codimension one subspace
$B$, so that $M$ splits into two parts $M_1$ and $M_2$. In this case we can do
the path-integral in two steps. Let us first fix the values of the fields
$\phi(x)$ to some given configuration $\phi'$ on $B$, and do the separate
integrals over $M_1$ and $M_2$ with these boundary conditions. If we write
\be
\Psi_{M_i}(\phi') = \!\!\Int_{\left.\phi\right|_B=\phi'}
\!\!\! [d\phi] \;e^{-S[\phi]}
\label{psi}
\ee
for the path-integral over the component $M_i$ with fixed values $\phi'$ at
$B$, it is clear that we will have
\be
Z(M) = \int [d\phi']\; \Psi_{M_1}(\phi') \cdot \Psi_{M_2}(\phi')
\ee
To make this more precise we have to consider general
space-times $M$ with boundaries.
To each connected component $B$ of the boundary $\d M$ of $M$ a
quantum field theory will associate a Hilbert space of states $\HH_B$
\be
B \ra \HH_B.
\label{HB}
\ee
In explicit examples this Hilbert space will be constructed by considering
canonical quantization on the cylinder $B \times \R$, where $B$ is considered
space-like and $\R$ the time direction. States $|\Psi\r \in \HH_B$ can be
represented as wave functions $\Psi(\phi)=\l\phi|\Psi\r$ of the fundamental
field variables $\phi(x)$ restricted to $B$. In our examples the vector spaces
$\HH_B$ will actually be finite-dimensional.

The assignment of vector spaces (\ref{HB})
should satisfy some intuitively clear axioms.
If the boundary $\d M$ has several disconnected components
$B_1,\ldots ,B_s$ we define
\be
\HH_{\d M} = \HH_{B_1} \otimes \cdots \otimes \HH_{B_s}.
\ee
Furthermore, if the boundary is empty, $\HH$ is one-dimensional
\be
\HH_\emptyset = \C.
\ee
The definition of $\HH_B$ will
in general depend on the orientation of $B$. If $-B$ denotes
the manifold with reversed orientation, we have
\be
\HH_{-B} = \HH^*_B,
\ee
since the path-integral on the cylinder $B \times \R$ gives a canonical
isomorphism
\be
\HH_B \otimes \HH_{-B} \ra \C.
\label{cylinder}
\ee
The space-time $M$ will defines a certain state
in the Hilbert space associated to its own boundary $\d M=B$
\be
|M\r\in \HH_B,
\ee
where the orientation of $M$ and $B$ agree.
Intuitively, the wave function $\Psi_M(\phi')$ representation of this state
is defined as the path-integral over all field configurations $\phi$ on $M$
that restrict to $\phi'$ at the boundary as in (\ref{psi}), see \fig 4.

In general we can consider the case that we give part of $\d M$ (say $B'$)
an orientation that agrees with that of $M$, and the other part ($B$)
the reverse orientation, as in \fig 5.
In that case the path-integral on $M$ will define an element in
$\HH_B^* \otimes \HH_{B'}$, or equivalently a transition amplitude
\be
\Phi_M \,:\ \HH_B \ra \HH_{B'}.
\ee
Factorization is now the property that if we cut $M$ in two parts
$M_1$ and $M_2$ along an intermediate slice $B''$, the transition amplitudes
satisfy
\be
\Phi_M = \Phi_{M_2} \circ \Phi_{M_1}.
\ee
This property is illustrated in \fig6. Since we compose linear operators, the
process includes a sum over states in the intermediate Hilbert space
$\HH_{B''}$.

We should stress that this type of factorization is characteristic of `matter
theories.' Theories of gravity, that somehow include an integral over all
metrics on $M$, behave characteristically differently. This is due to the fact
that in quantum gravity the path-integral only becomes well-defined if one
integrates over all metrics modulo diffeomorphisms. It is clear that not all
diffeomorphisms of a manifold $M$ respect the decomposition $M=M_1\cup M_2$, so
factorization of the path-integral is not obvious. Another important
modification in theories of quantum gravity is the natural and unavoidable
occurrence of singular configurations, where the topology of the manifold
actually changes. In Polyakov's word, the space of metrics has a boundary of
degenerate manifolds, and the path-integral naturally receives contributions of
this boundary. We met this phenomenon already when we discussed the stable
compactification of the moduli space of Riemann surfaces in \S2.1.

\newsubsection{Topological field theory in two dimensions}

In this subsection we will restrict our investigations to two dimensions
($D=2$) where our manifolds $M$ are surfaces of genus $g$. Since the only
connected compact one-dimensional manifold is the circle $S^1$, we have only
one vector space to consider
\be
\HH \equiv \HH_{S^1},
\ee
and for convenience we will assume it to be finite dimensional
\be
\dim \HH = N < \infty.
\ee
The data of a two-dimensional topological field theory are now obtained by
considering respectively the sphere with one, two, and three holes
\cite{thesis}. Let us briefly run through the argument.

{\it (i)} The disk gives rise to a particular state
\be
{\bf 1} \in \HH,
\ee
that we will denote as the identity, for reasons that become obvious in a
moment.

{\it (ii)} The cylinder gives a bilinear map
\be
\eta : \HH \otimes \HH \ra \C,
\ee
which we notate as
\be
\eta(a,b) = \l a, b \r.
\ee
By factorization, this inner product $\eta$ will be non-degenerate, but not
necessarily positive. It allows us to identify the `incoming states' in $\HH$
with the `outgoing states' in the dual space $\HH^*$. Note that here both
boundary components have an orientation that is opposite to that of the
cylinder, and the map $\eta$ should be sharply
distinguished from the identity map (\ref{cylinder}).

{\it (iii)} Finally the pair of pants,---or the sphere with three
holes---corresponds, again with the appropriate choice of orientations, to a
bilinear map
\be
c : \HH \otimes \HH \ra \HH,
\ee
If we introduce the notation
\be
c(a,b) = a \times b,
\ee
this makes $\HH$ into an algebra, the operator product algebra of the
topological field theory. Indeed, this allow us to identify states with
operators in a very simple way. It is convenient to choose an explicit basis
$\{\phi_0,\ldots,\phi_{N-1}\}$ for $\HH$, with $\phi_0={\bf 1}$, in terms of
which we have in component notation
\be
\eta_{ij}=\l \phi_i,\phi_j\r,\qquad
\phi_i\times\phi_j=\sum_k c_{ij}{}^k \phi_k.
\ee
All these data suffice to calculate any partition or correlation function,
since every surface can by factorization be reduced to a collection of
three-holed spheres. Of course, there are many inequivalent ways to factorize
a surface. The final answer should however not depend on the particular choice
of factorization. This concept, known as duality, gives further constraint on
the data $N$, $\eta$ and $c$. For instance, a simple consequence of the
symmetry of the 3-punctured sphere is the compatibility of the metric $\eta$
with the algebra $\HH$
\be
\l a \times b,c \r = \l a, b \times c \r.
\ee
If we consider the sphere with four holes, there are two inequivalent ways of
factorization as illustrated in \fig 7. This leads to so-called $s$-$t$
duality, which translates in associativity of the algebra $\HH$
\be
(a \times b) \times c = a \times (b \times c).
\ee
It can be easily checked that no further conditions will be found when we
consider more complicated surfaces.

With the concept of factorization, it is extremely easy to calculate
higher genus partition and correlation functions. In fact, we can introduce
an operator $H$ that creates an handle \cite{witten}. It is defined as
the state associated to the torus with one puncture (see \fig8), and has the
representation
\be
H = \! \sum_{a \in \rm basis}
\!\!(\Tr a^*) \cdot a = \sum_{i,j} c_i{}^{ij} \phi_j.
\ee
In this fashion a genus $g$ partition function $Z_g$ can be written as
a genus zero correlation function
\be
Z_g = \Bl \, \underbrace{H \cdots H}\subg \, \Br_0
\ee
{}From this expression it is immediately clear that for topological
field theories, in which the factorization axiom holds, partition functions
grow polynomial for large genus
\be
Z_g \sim c^g,
\ee
with some constant $c$. This behaviour contrasts starkly with results in
two-dimensional quantum gravity and string theory, where the size of the
partition function is related to the volume of moduli space and has the famous
factorial growth \cite{shenker}
\be
Z_g \sim (2g)!
\ee
This already indicates to us, that the simple kind of general covariance
that we are studying in this and the following section, is not the one relevant
for gravity theories. It has to be appropriately modified---we will have to
give up factorization---and we will return to these modifications in section 7.
However, before we discuss all this, we will first have to explain how
topological invariance can be realized in practice. Hereto we will
consider again the general $D$-dimensional case.

\newsubsection{The $Q$-cohomology}

It might not be obvious how all of the above structures can be realized in a
quantum field theory. Indeed, one does not expect in general such a simple
structure; finite dimensional Hilbert spaces are already exceptional in
quantum mechanics, let alone quantum field theory. The crucial ingredient is to
introduce a cohomology principle into the game.

The metric independence independence of a quantum field theory was
expressed by the vanishing of the energy-momentum tensor
\be
T_{\mu\nu} = 0.
\ee
In theories of quantum gravity this general covariance is achieved by making
the metric $g_{\mu\nu}$ a dynamic variable and integrating over all metrics in
the path-integral. The stress tensor of the matter is now balanced by the
contribution of the gravity sector
\be
T_{\mu\nu} = T_{\mu\nu}^{(matter)} + T_{\mu\nu}^{(gravity)} = 0.
\ee
However, the introduction of a fluctuating metric can be avoided if the quantum
field theory contains a fermionic charge $Q$ that is nilpotent
\be
Q^2 = 0.
\ee
One can think about $Q$ as a generating a BRST-like symmetry, like in gauge
theories, and define a physical Hilbert space $\HH$
as the cohomology ring of $Q$ in the full Hilbert space $\HH'$ of the original
field theory
\be
\HH = H^*_Q(\HH') = {\rm ker}\, Q /{\rm im}\, Q.
\ee
That is, physical states are annihilated by $Q$ and identified modulo
$Q$-exact states. It does not matter which particular operator we choose to
represent a class, since a well-known argument shows that spurious states
(states of the form $\{Q,\lambda\}$) decouple inside physical correlation
functions
\be
\l \{Q,\lambda\} \cO_1 \cdots \cO_s \r=0,\qquad \hbox{if $\{Q,\cO_i\}=0.$}
\ee
The bracket $\{\cdot,\cdot\}$ represents here either a commutator
or an anti-commutator, whichever is appropriate.
The existence of the operator $Q$ by itself is of course not enough to warrant
general covariance. To that end we need another essential ingredient,
namely that the energy-momentum tensor $T_{\mu\nu}$ is spurious itself
\be
\label{TQG}
T_{\mu \nu }=\{Q,G_{\mu \nu }\},
\ee
and thus cohomologically trivial.
The operator $G_{\mu\nu}$ must be a fermionic rank two tensor. The conserved
charges related to $T_{\mu \nu }$ and $G_{\mu \nu }$ are
\be
P_{\mu } = \int T_{\mu 0},\qquad
G_{\mu } = \int G_{\mu 0},
\ee
where the integrals are taken over a $D-1$ dimensional spacelike surface.
These charges form an anti-commuting extension of the translation group
\be
P_{\mu }=\{Q,G_{\mu }\}, \qquad \{G_{\mu },G_{\nu }\}=0,
\label{p}
\ee
that can be viewed as a supersymmetry algebra with charges of spin zero and
one, instead of the usual spin one-half.

\newsubsection{The descent equation}

Regarding the $Q$ symmetry as a spin zero supersymmetry is a very
fruitful analogy. In fact, it is convenient to go to a `superspace'
formulation of the theory, since this leads naturally to the
non-local observables in the model.
In addition to the previous space-time coordinates $x^{\mu}$, we will
now have $D$ Grassmannian coordinates $\theta ^{\mu}$. Starting from a
physical field $\phi(x)$, that we will sometimes denote as $\phi^{(0)}(x)$
to indicate that it is a local operator, the superfield $\Phi(x,\theta)$ is
given by
\ba
\Phi(x,\theta) \iss \exp{\theta^\mu G_\mu} \cdot \phi(x) \nonumber\\[3mm]
 \iss \phi^{(0)}(x) + \phi_\mu^{(1)}(x)\theta^\mu +
\ldots + \phi^{(D)}_{\mu_1 \ldots \mu_D}(x)\theta^{\mu_1}
\cdots \theta^{\mu_D}.
\label{superfield}
\ea
Here the fields $\phi ^{(k)}$ are generated from $\phi ^{(0)}$ by repeated
application of $G_{\mu}$
\be
\phi^{(k)}_{\mu_1 \ldots \mu_k}(x) = \{ G_{\mu_1},\{ G_{\mu_2},\ldots
, \{ G_{\mu_k},\phi ^{(0)}(x)\} \ldots \} \}.
\label{g}
\ee
Since $\phi^{(k)}_{\mu_1 \ldots \mu_k}$ is antisymmetric in all its indices, it
represents a $k$-form, and we can write (\ref{g}) symbolically as
\be
\{G,\phi^{(k)}\}= \phi^{(k+1)}.
\ee
Since by (\ref{p}) we have $\{Q,G\}=d$, these differential forms satisfy
the important {\it descent equation}
\be
d \phi^{(k)} = \{Q,\phi^{(k+1)}\}.
\label{descent}
\ee
We can draw two conclusions from this equation. First, it suggests a new class
of {\it non-local} physical observables. If $C$ is a $k$-dimensional
closed submanifold, the descent
equation shows that
\be
\phi(C) \equiv \int_{C} \! \phi^{(k)}(x)
\ee
is a physical observable, since
\be
\{Q,\phi(C)\} = \int_{C} \! d\phi^{(k-1)}
= \int_{\partial C} \! \phi^{(k-1)} = 0.
\ee
Secondly, if $C$ and $C'$ represent the same class in $H_k(M)$, we have
$C-C'=\d S$ and
\be
\phi(C)-\phi(C') =
\int_{S} d\phi^{(k)}= \{Q,\int_{S} \phi^{(k+1)}\}.
\ee
So the physical observable depends only on the homology class of $C$, as is
illustrated in \fig9.
That is, for each class in $H_k(M)$ and each element in $\HH$
we can construct a non-local operator.
In the case of the zero form fields, we have
\be
\phi^{(0)}(x) - \phi^{(0)}(y) =
\{Q,\int_x^y \! dx^{\mu} \,\phi^{(1)}_{\mu} \},
\ee
which shows explicitly our claim that the correlators of the local
physical operators do not depend on the positions.

\newsubsection{Perturbations of topological field theories}

One very important non-local operator, that always exists, is the top-form
\be
\phi(M) = \int_M\phi^{(D)}(x).
\label{volumeform}
\ee
The importance of thsi particular operators lies in the following fact.
Since $\phi^{(D)}$ is a volume form, it can be integrated over the space-time
itself. In this way we can make a modification of the action of the topological
quantum field theory by adding
(\ref{volumeform}) with some coupling coefficient $t$ to the action $S$
\be
S \ra S - t \cdot \int_M \phi^{(D)},
\ee
while preserving the general covariance. So, the top-dimensional partner of any
local observable defines a perturbation of the topological field theory. We can
in this way define a multi-parameter family of topological field theories whose
partition functions are given by
\be
Z(M,t) = \l \exp\sum_k t_k \cdot \int_M\phi_k^{(D)} \r.
\ee
The coupling coefficients $t_k$ can be seen as a moduli in the space of
topological field theories.

If we specialize to the case of two dimensions where $M$ is a surface $\Sigma$,
we see that apart from our
local observables
\be
\phi(P) \in \HH,
\ee
associated to a point $P\in\Sigma$, we also can consider the operators
\be
\phi(S^1)=\oint \phi^{(1)},\qquad \phi(\Sigma)=\int_\Sigma \phi^{(2)},
\ee
that correspond to a (non-trivial) cycle and the surface
$\Sigma$ itself. Temporarily discarding the one-forms, the most general
correlation function in the model is of the form
\be
\Bl \phi_{i_1}\cdots\phi_{i_s}\int\! \phi_{j_1}^{(2)}\cdots \int\!
\phi_{j_r}^{(
   2)}
\Br\subg
\ee
We have already given away how these correlators can in principle
be calculated. One simply considers the more general object
\be
\Bl \phi_{i_1}\cdots\phi_{i_s}\exp \sum_k t_k \int\!\phi_k^{(2)}\Br\subg
\ee
The expansion of this object in the coupling constants $t_k$ clearly generates
all possible inclusions of two-form operators. However, according to the above
reasoning it can alternatively be seen as a correlation function of only the
local operators $\phi_i$ in a new, perturbed topological field theory,
described by the coupling constants $t_k$. Since this new theory satisfies all
the axioms of \S5.3, all quantities can again be calculated by factorization.
That is, all essential information is encoded in the perturbed three-point
functions
\be
c_{ijk}(t)= \Bl \phi_i\phi_j\phi_k\exp \sum_k t_k \int\phi_k^{(2)}\Br\subzero
\ee
which define a family of multiplications on the vector space $\HH$.

So we finally arrive at the following beautiful picture of (a family of)
two-dimensional topological field theories. It is given by a bundle of
commutative, associative algebras $\HH$ fibered over a moduli space $X$. The
data $N$, $\eta$, and $c$ vary continuously and satisfy the relations of
\S5.3. Moreover, there is a well-defined map from a state $\phi_k \in \HH$ to a
vector field $\d/\d t_k$ on $X$. Furthermore, these vector fields commute, and
can be integrated to give local coordinates $t_k$. In the following section we
will give some very explicit examples of all this.

\newsection{Topological Conformal Field Theory}

We have seen that a general two dimensional topological field theory is solved
in terms of the perturbed three-point functions $c_{ijk}(t)$. Our aim will now
be to calculate these quantities in some examples. To this end we have to
introduce some auxiliary structure: we will impose the condition that the
topological field theory be conformally invariant. This might seem a bit
superfluous. All {\it physical} quantities in a topological field theory
are automatically invariant under conformal transformations, since they are
independent of the metric. However, we demand that even the non-physical
sector---the `big' Hilbert space $\HH'$---be conformally invariant. We will
call these models {\it topological conformal field theories} (TCFT), see also
\cite{d<1,notes}.

\newsubsection{Twisting $N=2$ superconformal models}

Conformal invariance requires that the trace of the stress-energy tensor
vanishes
\be
T_{\mu}{}^{\mu} = 0,
\ee
not just within physical correlation functions, but identically as an operator.
The stress tensor still has a $Q$ partner $G_{\mu\nu}$
\be
T_{\mu \nu} = \{Q, G_{\mu \nu}\},
\ee
which is also assumed to be traceless
\be
G_{\mu}{}^\mu = 0.
\ee
In complex coordinates $(z,\zbar)$ the two remaining components $T=T_{zz}$ and
$\Tbar=T_{\zzbar\zzbar}$ are respectively holomorphic and anti-holomorphic
fields. We will for the moment restrict our attention only to the left-moving
fields, \ie, fields holomorphic in $z$. Let us make a list of the different
fields that appear in this sector.

First of all we have the stress tensor $T(z)\,dz^2$ and its $Q$-partner
$G(z)\,dz^2$, both quadratic differentials. These have the standard
decompositions into modes
\be
T(z) = \sum L_n z^{-n-2},\qquad G(z)= \sum G_n z^{-n-2}.
\ee
We will make the further important assumption, that the Hilbert space $\HH'$
will decompose in highest weight modules of the algebra generated by the $L_n$
and $G_n$, which is an extension of the $c=0$ Virasoro algebra
\be
[L_m,L_n] = (m-n)L_{m+n},
\ee
by a vector supersymmetry:
\be
[L_m,G_n] = (m-n)G_{m+n},\qquad \{G_m,G_n\}=0.
\ee
The conformal central charge $c$, that in principle could appear in the
Virasoro algebra, must necessarily vanish, since a conformal anomaly is not
allowed in a general covariant theory, \cf\ critical string theory. We have
primary fields $\phi_i$ satisfying
\be
L_n|\phi_i\r=0,\qquad G_n|\phi_i\r=0,\quad(n>0).
\ee
Notice the second relation, which is not a consequence of the first. Actually
in this respect conformal invariance is somewhat of a misnomer, since we really
require invariance under the larger algebra generated by both the stress tensor
$T$ and its anti-commuting partner $G$.

There is a second multiplet of fields related to the (anti-commuting)
BRST-current $\Q(z)\,dz$, whose conserved charge is the left-moving part of the
scalar $Q$-operator
\be
Q=Q_0 + \Qbar_0,
\ee
with
\be
Q_0 = \oint \Q(z)\,dz,\qquad \Q(z) = \sum Q_n z^{-n-1}.
\ee
We will use the notations $Q$ and $Q_0$ interchangeably if no confusion arises.
Since the BRST-current is a spin one field, it is---by an argument that we will
sketch in a moment---necessarily exact:
\be
\Q(z) = \{Q,J(z)\},\qquad J(z) = \sum J_n z^{-n-1},
\ee
where $J(z)\,dz$ is a bosonic field of spin one, that is, a honest abelian
current. So we arrive at the multiplet of holomorphic fields listed in table 1.
\tabel{
\begin{center}
\renewcommand{\arraystretch}{1.5}
\begin{tabular}{||c||c|c|c|c||}
\hline
\strut\hbox{current} & \hbox{spin $h$} & \hbox{charge $q$} &
$h+\half q$ & \hbox{statistics} \\ \hline
$T(z)$ & 2 & 0 & 2 & {\rm boson} \\
$G(z)$ & 2 & -1 & ${3\over2}$ & {\rm fermion} \\
$Q(z)$ & 1 & 1 & ${3\over2}$ & {\rm fermion} \\
$J(z)$ & 1 & 0 & 1 & {\rm boson} \\
\hline
\end{tabular}
\renewcommand{\arraystretch}{1.0}
\end{center}}{The conserved currents of a topological conformal field
theory with their conformal dimensions, charges, and statistics.}

All fields $\phi_i(z,\zbar)$ in the theory can be chosen to
have some definite conformal dimensions $h_i$ and charges $q_i$
\be
L_0 |\phi_i\r= h_i |\phi_i\r,\qquad
J_0 |\phi_i\r = q_i |\phi_i\r.
\ee
Since we have
\be
L_0=\{Q,G_0\},\qquad [Q,L_0]=0,
\ee
we see immediately that all physical states, satisfying $Q|\phi\r=0$,
are ground states, with $h=0$.
Indeed, suppose $L_0|\phi\r=h|\phi\r$ with $h\neq 0$, then by a familiar
argument this state is exact
\be
|\phi\r={1\over h} L_0 |\phi\r= Q|\lambda\r,\qquad
|\lambda\r={1\over h}G_0|\phi\r.
\ee

The multiplet $\{T,G,Q,J\}$ is very reminiscent of the generators of the $N=2$
superconformal algebra, where we would have a stress tensor $T$, two spin $3/2$
supercurrents $Q^+, Q^-$ and a $U(1)$ current $J$. The two sets of currents are
related by the `twist' \cite{tft,sigma,E-Y}
\be
T \ra T + \half \d J,
\label{twist}
\ee
which gives in particular $L_0 \ra L_0 + \half J_0$ and accordingly adds the
charge to the conformal dimensions (see table 1)
\be
h \ra h + \half q.
\ee
We can now consider the algebra generated by the modes $L_n,G_n,Q_n,J_n$ of
the four currents. If they form a {\it closed} algebra---which need not be the
case, of course---their commutators are completely fixed, due to the uniqueness
of the $N=2$ superconformal algebra, that we would obtain after twisting. In
this way we arrive at the following algebra

\begin{eqnarray}
\label{TCFTalg}
\lefteqn{
\begin{array}{ll}
 [L_m,L_n]\; =\; (m-n)L_{m+n},\ \ \ \ & \ \ \ \ \
 [J_m,J_n]\; =\; d\cdot m \,\delta_{n+m,0},\\[4mm]
[L_m,G_n]\; =\; (m-n)G_{m+n},\ \ \ \ \ & \ \ \ \ \
 [J_m,G_n]\; =\; -G_{m+n},\\[4mm]
[L_m,Q_n]\; =\; -n\,Q_{m+n},\ \ \ \ & \ \ \ \ \
 [J_m,Q_n]\; =\; Q_{m+n},
\end{array}}
\nonumber\\[4mm]
& \begin{array}{ll}
\ \ \{G_m,Q_n\} & \!\!\!=\; L_{m+n}+nJ_{m+n}+
{\textstyle{1\over 2}}d\cdot m(m\! +\! 1) \delta_{n+m,0}, \\[4mm]
\ \ \, [L_m,J_n]\, & \!\!\!=\; -n J_{m+n}
-{\textstyle {1\over 2}}d\cdot m(m\! +\! 1)\delta_{m+n,0}.
\end{array} &
\end{eqnarray}

\noindent Most of these relations are self-evident; they simply express the
conformal dimensions and charges of the currents. We note the appearance of a
central charge $d$ for the $U(1)$ symmetry. In the case of a topological
sigma-model $d$ will correspond to the complex dimension of the target space,
which explains the choice of symbol. The same constant features as an anomaly
in the commutator of the stress tensor with the $U(1)$ current; we will return
to this point momentarily. The central charge for the corresponding $N=2$
supersymmetric theory is related to the anomaly in the topological model by
$c=3d$.

The anomaly in the current $J(z)$ leads to selection rules for the correlation
functions. $J(z)$ does not transform as a proper current under coordinate
transformations, as we read off from the commutation relations of $L_n$ with
$J_m$ in (\ref{TCFTalg}). If $z \ra w(z)$ denotes a holomorphic coordinate
transformation, then
\be
J_0 \ra J_0 + d \cdot \oint \d_z (\log \d_z w),
\ee
where the additional term measures the winding number of the map $\d_z w$.
So, under the inversion $w =1/z$ the zero mode $J_0$ goes to
$J_0-d$. This results in a background charge of $-d$ on the sphere.
The background charge for arbitrary topology can be found from the covariant
expression of the $U(1)$ anomaly
\be
\nabla_{\mu}J ^{\mu} =-{d\over 8\pi} \sqrt{h}\, R,
\ee
where $h_{\mu \nu}$ is the world-sheet metric and $R$ is the associated
scalar curvature. Consequently, since the curvature density integrates
to the Euler character $\chi=2-2g$ on a genus $g$ surface,
there is a background charge of $d \cdot (g-1)$ present that has to
be compensated. Correlation functions
\be
\l \phi_{i_1} \ldots \phi_{i_s} \r\subg
\ee
therefore obey the selection rule
\be
\sum_{i=1}^s q_i = d(1-g).
\ee
In relation to this we note that the two-form operators
\be
\phi_i^{(2)} = G_{-1} \Gbar_{-1} \phi_i^{(0)}
\ee
have $(\hbox{\it left},\hbox{\it right})$ charges $(q_i -1, \qbar_i -1)$.

By the above twisting {\it any} $N=2$ superconformal model produces a
topological CFT. This not only provides a wealth of examples, but also allows
us to use some very interesting results derived in the context of $N=2$
supersymmetry. For instance, the unitary $N=2$ models come with a natural
positive inner product, that we can carry over to the topological field theory.
Because of the twisting, this inner product is not Lorentz-invariant. In fact,
with this inner product we have the hermiticity property $Q^\dagger = G_0$,
which relates a scalar and a vector quantity. By considering the relation
\be
0=\l\phi|L_0|\phi\r=\|Q|\phi\r\|^2+\|G_0|\phi\r\|^2,
\ee
one shows that in this case the ground states satisfy both
\be
Q |\phi\rangle=0,\qquad\qquad G_0|\phi\rangle = 0.
\ee
The latter condition picks out a unique representative for each $Q$-cohomology
class. It should be compared to harmonic forms in ordinary de Rham cohomology
theory. In the $N=2$ context these fields are known as chiral primary fields
\cite{LVW}.

\newsubsection{Generalized $SL(2,C)$ invariance and its consequences}

We now want to derive an important symmetry property of topological conformal
field theories. To this end we recall that the non-local operators can be
combined into one superfield (\ref{superfield})
\be
\Phi(\theta,\thetabar) = \phi^{(0)} + \phi^{(1,0)}\theta
+ \phi^{(0,1)}\thetabar + \phi^{(2)}\theta\, \thetabar.
\ee
Here we suppressed the $(z,\zbar)$ dependence. Let us now consider a
correlation function on the sphere of the form
\be
\Bl \prod_{i=1}^s \int d^2\!z\,d^2\theta\; \Phi_i(z,\zbar,\theta,\thetabar)
\Br\subzero.
\ee
This expression is not well-defined, since it is invariant under a fermionic
extension of the well-known $SL(2,C)$ symmetry, generated by operators $L_0$,
$L_1$, $L_{-1}$ and $G_0$, $G_1$, $G_{-1}$. We have to factor out the infinite
volume of this group in order to obtain a finite answer.
These symmetries correspond to the
super-M\"obius transformations
\be
z \rightarrow {az+b\over cz+d},\qquad \theta \rightarrow
{\theta + \alpha z^2+\beta z+\gamma\over (cz+d)^2}.
\ee
This extended $SL(2,C)$ symmetry can be used to fix three of the
$z$-coordinates, say $z_1,z_2,z_3$, at $0$, $1$, and $\infty$, and put three of
the $\theta$-coordinates to zero. We choose these anti-commuting coordinates to
be $\theta_1,\theta_2,\theta_3$. If we recall that
\be
\int d^2\theta\cdot\Phi = \phi^{(2)},\qquad
\left. \Phi \right|_{\strut \theta=0} = \phi^{(0)},
\ee
then we see that after gauge fixing we are left with a correlation function of
the form
\be
\label{Widentity}
\Bl\phi^{(0)}_{i_1}\phi^{(0)}_{i_2}\phi^{(0)}_{i_3} \int\!
\phi^{(2)}_{i_4}\ldots \int\phi^{(2)}_{i_s}\Br\subzero.
\ee
Since we started from an expression that was explicitly symmetric in {\it all}
indices $i_1,\ldots,i_s$, this correlator also has this permutational symmetry.
That is, it does not matter which three operators we represent as
zero-forms---the generalized $SL(2,C)$ invariance tells us that we can
interchange a zero and a two-form. An alternative derivation of this result
uses the Ward identities associated to this global invariance \cite{d<1}.

In terms of the coefficients $c_{ijk}(t)$ the permutation symmetry of
(\ref{Widentity}) gives the important integrability condition
\be
{\partial c_{ijk} \over \partial t_l} =
{\partial c_{ijl} \over \partial t_k}.
\label{4pt}
\ee
We stress that this is an additional condition imposed on the family of
algebras $c(t)$, and a consequence of the topological invariance. We can
integrate this relation three times---at least locally---to find that the
three-point functions are actually the third derivatives of a function $F(t)$,
the so-called free energy
\be
\label{dddf}
c_{ijk}(t) = {\partial^3 F(t) \over \partial t_i \partial t_j \partial t_k}.
\ee
Symbolically, $F(t)$ is defined as
\be
F(t) = \Bl \exp \sum_i t_i\!\!\int\!\Phi_i\Br,
\label{free}
\ee
that is, $F=\log \tau$, with $\tau(t)$ the string partition function.

As a corollary to the result (\ref{4pt}), consider the special identity
operator $\phi_0 = {\bf 1}$. There is no corresponding two-form, since $G$
commutes with the identity. So the coupling coefficient $t_0$ does not
exist---there is no modulus associated to ${\bf 1}\in \HH$. This fact, combined
with integrability relation (\ref{4pt}), shows that
\be
0 = {\partial c_{ijk} \over \partial t_0} =
{\partial c_{ij0} \over \partial t_k} = {\partial \eta_{ij} \over \partial
t_k}.
\label{metric2}
\ee
So we have shown that the metric or two-point function $\eta_{ij}$ is
independent of the deformation parameters $t_k$.

Summarizing, we have derived in this section three additional important
results, that followed from the conformal invariance. First, all of the
coefficients $c_{ijk}(t)$ are derived from a single function, the free energy
$F(t)$. This function is well-defined on local patches of the moduli space of
TCFT's. Secondly, the metric $\eta_{ij}(t)$ on the Hilbert space of states is
in
fact independent of the couplings $t_{k}$. This is equivalent to the statement
that $\d F/\d t_0$ is a quadratic function. Finally, there exists an anomalous
$U(1)$ symmetry, which assigns charge $q_i$ to the local physical fields
$\phi_i$, with a background charge $d$, and which corresponds to a scaling
relation for the free energy of the form
\be
\label{scaling} \sum_j(q_j-1)t_j\ddt{j} F(t)=(d-3)F(t).
\ee
Remarkably enough, these three ingredients will be sufficient to solve the
theory in many special cases, in particular for all models with $d<1$
\cite{d<1}.

\newsubsection{Topological Landau-Ginzburg models}

At this point it might be helpful to consider some examples of TCFT's. The
simplest example is a free field theory of $d$ complex bosons and fermions. It
can obtained by twisting $d$ copies of the $c=3$ free field realization of the
$N=2$ superconformal algebra. The bosons will be written as $x^i,x^\ibar$,
with $i=1,\ldots,d$. In order to describe the fermions we have to make a
choice. This choice is related to the following ambiguity. Up to now we
basically only discussed the left-moving sector. Here the topological and the
superconformal model were related by the twist
\be
T \ra T \pm \half \d J.
\ee
Here the $\pm$ sign is completely irrelevant, since we have no absolute way to
fix the overall sign of the $U(1)$ charge. However, if we combine left and
right-movers we can consider either the $(+,+)$ or the $(+,-)$ version.
Although the resulting TCFT's are still equivalent, since {\it
quantum-mechanically} we can also not distinguish the signs of the separate
left and right-moving currents, the interpretation in terms of the {\it
classical} geometry of the original superconformal model might be vastly
different. In all its generality this phenomenon is known as `mirror symmetry'
\cite{mirror}. Even in case of our simple free field theory we can consider two
different actions and symmetry realizations that lead to equivalent quantum
field theories. The problem manifests itself here only in the spin assignments
of the fermions. In the $N=2$ model we have four different kinds of fermionic
operators that we can denote as
\be
\psi_L^i(z),\ \psi_R^i(\zbar),\ \psi_L^\ibar(z),\ \psi_R^\ibar(\zbar).
\ee
These fields, all of spin one-half, obtain the unusual spins of zero and one
after twisting. The two possible twists identify as the spin zero fields either
the pair $\psi_L^i, \psi_R^\ibar$ or the pair $\psi_L^i, \psi_R^i$. We will
continue to work with the latter choice, and will combine the two fields into
one field
\be
\psi^i(z,\zbar) = \psi_L^i(z) + \psi_R^i(\zbar).
\ee
We denote the spin one fields as $\chi^\ibar_z, \chi^\ibar_\zbar$. They are the
two components of a one-form $\chi^\ibar_\mu$. The action of these fields
reads, with a summation over $i=\ibar$ understood,
\be
S_0 = \int d^2\!z
\left(\d_\mu x^\ibar \d^\mu x^i + \chi_\mu^\ibar \d^\mu \psi^i\right).
\ee
The next step is to identify the nilpotent charge $Q$. With
$\delta=\{Q,\cdot\}$ the non-vanishing transformations are simply
\be
\delta x^i= \psi^i,\qquad \delta \chi_\mu^\ibar= -\d_\mu x^\ibar.
\ee
With these field transformations we easily verify that the action is invariant.
The stress-energy tensor is indeed $Q$-exact
\be
T(z) = \d_z x^\ibar \d_z x^i + \chi^\ibar_z\d_z\psi_L^i = \{ Q, G\},
\qquad G(z)=-\chi_z^\ibar \d_z x^i.
\ee
The set of currents is complemented in this example by
\be
Q(z) = \d_z x^\ibar\psi_L^i,\qquad J(z) = -\chi_z^\ibar\psi^i.
\ee
This trivial free field theory can be generalized in two interesting
directions. First, we can consider non-linear sigma models, where the flat
space-time of one-complex dimension is replaced by an arbitrary $d$-dimensional
Ricci-flat K\"ahler manifold $M$. In superfield notation, with $X^i=x^i +
\theta \psi^i + \ldots$, the action reads
\be
S = \int d^2\!z\,d^4\theta\cdot K(X^i,X^\ibar).
\ee
where $K$ is the K\"ahler potential, and $x^i,x^\ibar$ are the local
coordinates on $M$. These models are well-known to possess $N=2$ superconformal
invariance, which translates into topological invariance after twisting. In the
above model we simply have chosen the flat target space metric
\be
K(X^i,X^\ibar)= X^i X^\ibar.
\ee
\newcommand{\Wbar}{\overline{W}}
A second possible generalization is the inclusion of a superpotential $W(X)$.
In this case we have a so-called topological Landau-Ginzburg model \cite{vafa},
see also \cite{ito}. The action reads
\be
S = S_0 + \int d^2\!z\,d^2\theta\cdot W(X^i) + c.c.
\ee
or, after the integration over $(\theta,\thetabar)$ and eliminating the
auxiliary fields,
$$
S = \int d^2\!z\;\Bigl(\d_z x^\ibar \d_\zzbar x^i + \chi_z^\ibar
\d_\zzbar \psi_L^i + \chi_\zzbar^\ibar \d_z \psi_R^i \qquad\qquad\qquad\quad
$$
\be
\qquad\qquad + \ \Dd {x^\ibar} W \Dd {x^i}{\Wbar}
+ \psi_L^i\psi_R^j{\d^2 W \over \d x^i\d x^j} + \chi_z^\ibar\chi_\zzbar^\jbar
{\d^2 \Wbar \over \d x^\ibar \d x^\jbar}\Bigr).
\ee
In this case we have to modify the transformation rules by \cite{vafa}
\be
\delta \psi^i_L = \Dd {x^\ibar} W,\qquad
\delta \psi^i_R = -\Dd {x^\ibar} W.
\ee
Landau-Ginzburg models become only
conformal invariant at the renormalization group fixed point, if the
superpotential is quasi-homogeneous \cite{mart,VW}. That is, if there
exists a scaling law, with charges $q_i$, such that (from now we will use
lower index notation $x_i\equiv x^\ibar$ to avoid confusion with powers of
the coordinates)
\be
W(\lambda^{q_i}x_i) = \lambda W(x_i).
\ee
Landau-Ginzburg models have the wonderful property that the correlation
functions of the chiral primary fields---or, after twisting, the physical
fields--- are completely determined by the superpotential $W$. In fact, for a
quasi-homogeneous potential $W(x_i)$ the local physical fields are simply
functions $f(x_i)$ of the invariant field $x_i\equiv x^\ibar$. Since we have
\be
{\d W \over \d x_i} = \{Q,\psi_L^i\},
\ee
every function of the form $f=g^i \d_iW$ is $Q$-exact, and the physical
state space is accordingly given by
\be
\HH ={\C[x] \over \d W(x)}.
\label{ring}
\ee
Since there are no singularities in the operator product of two function
$f(x_i)$, the operator algebra does not receive corrections, and is also given
by (\ref{ring}) now considered as a polynomial ring, where $\C[x]$ is the
set of polynomials in the $x_i$, and we have factored out the ideal generated
by
\be
\Dd {x_i} W = 0.
\ee
The charge of an element $\phi(x)$ is determined by assigning charge $q_i$ to
the fundamental field $x_i$.

In general we can consider deformations of $W$ by an element in $\HH$ that
preserve the quasi-homogeneity, \ie, an operator of charge $q=1$, a modulus or
marginal operator. A simple example is the potential
\be
W=x^3+y^3+z^3+a\cdot xyz.
\ee
This clearly introduces a modulus---for any value of $a$ the potential is
homogeneous, whereas the different rings $\HH$ are non-isomorphic. There is a
special set of potentials that have no moduli. The corresponding field
theories are known as topological minimal models---the twisted versions of
the minimal $N=2$ superconformal models, and as such were first considered by
Eguchi and Yang \cite{E-Y} and, in relation with solvable string theories, by
Li \cite{Li},. In this way the classification of rational singularities enters
the subject, since these potentials have an $ADE$ classification, as given
in table 2. This is of course the same $ADE$ labeling that features in the
classification of modular invariant partition functions in minimal conformal
field theories \cite{ADE}.
\tabel{
\vspace{-10mm}
\begin{center}
\small
\begin{tabular}{||c||c|c|c||} \hline
\bliep $G$ & $\Gamma$ & $W$ & $d$ \\ \hline
\bliep $A_n$ & \An & $x^{n+1}$ & ${n-1\over n+1}$\\ \hline
\bliep $D_n$ & \Dn & $x^{n-1} + xy^2$ & ${n-2\over n-1}$\\ \hline
\bliep $E_6$ & \Esix & $x^3+y^4$ & ${4\over5}$ \\ \hline
\bliep $E_7$ & \Eseven & $x^3+xy^3$ & ${8\over 9}$ \\ \hline
\bliep $E_8$ & \Eeight & $x^3+y^5$ & ${14\over15}$ \\ \hline
\end{tabular}
\end{center}}{The simply-laced Lie groups $G$, their Dynkin diagrams $\Gamma$,
the corresponding superpotentials $W$, and the central charge $d$ of
the topological minimal model.}

It is not difficult to give the explicit forms of the rings $\HH$.
For instance, the $A_{p-1}$ ring with
\be
W(x)=x^p
\ee
is simply given by
\be
x^{p-1}=0.
\ee
Thus, a basis is furnished by the elements
\be
\left\{ 1,x,x^2,\ldots , x^{p-2} \right\}.
\label{Aring}
\ee
Quite generally the fields $\phi_i\in\HH$ are in one-to-one correspondence with
the exponents of the simply-laced Lie group $G$. The field $\phi_i$ has charge
$q_i = i/p$ with $i+1$ an exponent, and $p=h_G$, the dual Coxeter number of
$G$. The `central charge' $d$ of the minimal model is given by
\be
d={p-2\over p},
\ee
see also table 2. The metric $\eta$ is determined through charge conservation,
and is of the skew-diagonal form
\be
\eta_{ij}=\delta_{i+j,p-2}.
\ee

\newsubsection{The perturbed operator algebra}

It is clear that a choice of superpotential $W(x)$ determines the operator
algebra (\ref{ring}) and therefore all correlation functions of the local
operators $\phi^{(0)}$. As we have explained, by factorization, it suffice to
calculate the perturbed three-point function
\be
c_{ijk}(t) = \Bigl\langle\phi_i\phi_j\phi_k
\exp\sum_nt_n\!\int\!\phi^{(2)}_n\Bigr\rangle_0
\label{cijkt2}
\ee
in order to include also the non-local two form $\phi^{(2)}$. We will be able
to calculate this algebra as a result of the existence of a flat metric on the
moduli space of TCFT's. Much of the structure we will meet, in particular the
existence of `flat coordinates' associated to a rational singularity, has
already appeared in the mathematical literature through the work of K. Saito
\cite{saito}. Recently, these techniques have been applied in the context of
topological field theory \cite{blok,lerche}. We will sketch here the less
sophisticated argument of \cite{d<1}.

For definiteness we will consider only the models in the $A$-series, so that
$W(x) = x^p,$ and the basis elements of $\HH$ are given by
\be
\phi_i = x^i,\qquad\quad i=0,1,\ldots,p-2.
\ee
The crucial idea is that the perturbed algebra is still of the form
(\ref{ring}), but with a more general superpotential
\be
W= x^p + \sum_{i=0}^{p-2} g_i(t)x^i.
\label{pertW}
\ee
Intuitively, a result of this form is to be expected, since we add to the
action
the operators
\be
\phi^{(2)}_i = \int d^2\theta \cdot X^i.
\ee
However, the extra terms on the RHS of (\ref{pertW}) are not of the simple form
$t_i x^i$. This is the case because the operators are only given by $\phi_i =
x^i$ at the conformal ($t=0$) point, and become more complicated $t$-dependent
polynomials after a general perturbation. We therefore should allow for more
general functions $g_i(t)$ that behave as $g_i=t_i+\OO (t^2)$. For general
values of the coupling constants the fields $\phi_i$ are defined by
\be
\phi_i(x,t)= \Ddt i W = \sum_{j=0}^{p-2} {\d g_j \over \d t_i} x^j.
\label{def1}
\ee
We want to calculate the functions $g_i(t)$, or equivalently the fields
$\phi_i(x,t)$.

Of course, one can also consider the $g_i$'s as coordinates on the space of
topological field theories, and this is the perfect legitimate point of view
taken in \cite{vafa}. In that case one should worry about contact terms that
are quite generally associated to reparametrizations of moduli \cite{kutasov}.
The preferred coordinates $t_k$, that naturally follow from the conformal
invariance underlying the model, are referred to as `flat coordinates.' The
reason for this---not quite appropriate---nomenclature is the following. In
\S6.2 we proved that in a general topological conformal field theory the metric
\be
\eta_{ij} = \l \phi_i \phi_j \r\subzero
\ee
is constant, when consider a function of the moduli $t_i$. That is, we always
have a flat metric on the space of perturbed TCFT's and we can choose
coordinates $t_i$ such that this metric is constant. We will use this crucial
ingredient to determine the correlation functions. Notice that for a {\it
given} superpotential $W$ the operator products are completely
determined---they are given by the polynomial ring
\be
\phi_i(x)\,\phi_j(x) = \sum_k c_{ij}{}^k \phi_k(x) \qquad ({\rm mod}\ W\p(x)).
\label{OPEW}
\ee
The inner product should be compatible with this multiplication, and is
thereby uniquely determined to be
\be
\l \phi_i \phi_j \r = \oint {dx \over 2\pi i} \,
{\phi_i(x)\, \phi_j(x) \over W\p (x)}.
\ee
(Here we used in an essential way the fact that $W$ allows no moduli. Otherwise
this metric would only be determined up to a conformal factor. See for instance
\cite{verlinde-warner} for an example where this phenomenon occurs.) The above
metric clearly respects the operator algebra (\ref{OPEW}), since terms
proportional to $W'(x)$ do not contribute. It can also be directly derived from
the path-integral representation \cite{vafa}. We know that the fields $\phi_i$
should be orthogonal with respect to this inner product. These orthogonal
polynomials can be defined in the following elegant way as derivatives of
fractional powers of the superpotential
\be
\phi_{i}(x) = {1\over i+1}{d W_+^{(i+1)/p}}.
\label{def2}
\ee
Here $d=\d/\d x$, and the $+$ indicates again a truncation to positive powers
of $x$, similar as in our discussion of the KdV hierarchy in \S3.1. One easily
verifies, with $z=W^{1/p}$ that indeed
\ba
\l\phi_i \phi_j\r \iss {1\over2\pi i} \oint {(z^idz)_+ \, (z^j dz)_+
\over z^{p-1} dz}\nonu
\iss {1\over2\pi i}\oint z^{i+j-p+1} dz = \delta_{i+j,p-2}.
\ea
The reader is encouraged to verify that the $+$'s have legitimately been
dropped in this derivation. The solution of the model results if we combine the
equations (\ref{def1}) and (\ref{def2}), and obtain the important result
\be
(i+1)\Ddt i W = \dd x {W^{(i+1)/p}_+}.
\label{dW}
\ee
After a relabelling $t_i \ra (i+1)t_{i+1}$ this is exactly the equation
(\ref{top-kdv}) that described the first few `primary' flows in the $p\th$ KdV
hierarchy. We can identify the fundamental Landau-Ginzburg field $x$ with the
classical momentum $y=D$, and the superpotential $W(x)$ with the classical
version of the Lax operator $L(y)$. In this way the operators are related by
\be
\cO_k = k\cdot \phi_{k-1}=dW^{k/p}_+.
\ee
Some more work is needed to show that also the string partition function and
the $\tau$-function coincide. In this way we have recovered a crucial part of
the quantum field theory underlying the structures of \S\S2--4. The twisted
$N=2$ minimal models where first proposed as the relevant quantum field
theories for the $(p,1)$ string theory by Keke Li \cite{Li}. As explained in
\cite{Nmatrix,milnor} these minimal models naturally give rise to the top Chern
classes that appeared in the intersection numbers defined in \S2.4.

In case the reader is wondering: equation (\ref{dW}) is sufficient to
explicitly solve the operators $\phi_i(x)$ and the superpotential $W(x)$. The
result reads
\be
\phi_i(x) = (-1)^i \det \left(
\baccccc
-x & 1 & 0 & \cdots & 0 \\
t_{p-2} & -x & 1 & \ddots & \vdots \\
t_{p-3} & t_{p-2} & \ddots & ~ & 0 \\
\vdots & ~ & \ddots & ~ & 1 \\
t_{p-i} & \cdots & t_{p-3} & t_{p-2} & -x
\eac \right)
\ee
The superpotential is obtained by putting $\d W/\d x=\phi_{p-1}$ in the above
equation.

\newsection{Topological String Theory}

In this final section we will start our discussion of topological gravity and
topological string theory. Through these vastly more complicated quantum field
theories we will be able to make contact with the intersection theory and
integrable hierarchies that featured in the first few sections.

There exists a general recipe to associate to a given moduli space
a topological quentum field theory whose correlation functions reduce to
intersection numbers of cohomology classes of the moduli space
\cite{witten-trieste}. In the case of the moduli space of Riemann surfaces this
field theory is known as topological gravity \cite{topgrav}. The fundamental
multiplet consists of the spin connection $\omega_\mu$, its anti-commuting
partner $\psi_\mu$, and a bosonic scalar field $\phi$. The BRST transformations
read
\be
\delta \omega_\mu=\psi_\mu,\qquad
\delta\psi_\mu=\d_\mu\phi,\qquad
\delta\phi=0.
\ee
The cohomology classes $\sigma_n$ correspond to the physical observables
\be
\sigma_n = \phi^n.
\ee
The relation with intersection theory is in this formulation conceptually most
clear. However, it does not lead easily to explicit calculations. Therefore we
choose here a slightly more complicated, but equivalent, representation due to
Erik and Herman Verlinde \cite{verlinde,notes}. We will refer to this theory
in all its generality as {\it topological string theory.}

As in conventional string theory, topological strings will consist of a matter
theory coupled to Liouville and ghost fields. The matter sector will be a
topological CFT as was studied in the previous section. We have seen that the
existence of the nilpotent $Q$-charge simplified things enormously. Our aim
will be to introduce a dynamical metric, or, after gauge fixing, a
Liouville-like field and ghosts, while preserving the $Q$-symmetry at all
stages. The fundamental symmetries, generated by the stress-tensor $T_{\mu\nu}$
and its partner $G_{\mu\nu}$, that we encountered in the previous section, will
now be lifted to the status of gauge symmetries. In our presentation we will
follow very closely \cite{verlinde,notes}. We will explain the construction
starting with our most simple example.

\newsubsection{Action and symmetries}

Let us reconsider the $d=1$ topological sigma model of \S6.3 whose action we
now write for a general world-sheet metric $g_{\mu\nu}$ as
\be
S = \int d^2\!z\;
\sqrt g g^{\mu \nu} (\d_{\mu} x \d_{\nu} \xbar +
\chi_{\mu} \d_{\nu} \psi ).
\label{minLag}
\ee
We want to couple this model
to gravity, so that the metric $g_{\mu\nu}$ becomes
dynamical, without ruining the topological invariance guaranteed by
the $Q$-symmetry. Therefore, each field must have a fermionic partner.
So, we introduce the metric's fermionic partner $\psi_{\mu\nu}$ together
with the transformation rules
\be
\delta g_{\mu \nu} = \psi_{\mu \nu}, ~~~~~~~~~~~
\delta \psi_{\mu \nu} = 0.
\ee
Just as variations of the action with
respect to the metric give the stress-energy tensor, variations with
respect to $\psi_{\mu \nu}$ give its partner $G_{\mu \nu}$
\be
T^{\mu \nu} = {1\over \sqrt g}\frac{\delta S}{\delta g_{\mu \nu}},\qquad
G^{\mu \nu} = {1\over \sqrt g}\frac{\delta S}{\delta \psi_{\mu \nu}}.
\label{actionvars}
\ee
We already found that $G_{\mu\nu}$ was given by
$\chi_\mu\d_\nu \psi$.
These functional derivatives can be integrated to get the full action
describing the coupling of topological matter to topological gravity
\be
S_{m}= \int d^2\!z\; \sqrt g g^{\mu \nu} (\d_{\mu} x \d_{\nu} \xbar +
\chi_\mu \d_\nu \psi + \psi_{\mu \rho} \chi ^{\rho}\d_\nu x ).
\label{fullact}
\ee
We will work in the conformal gauge
\be
g_{\mu \nu} = e^{\phi} \delta_{\mu \nu},\qquad
\psi_{\mu \nu} = \rho \,e^{\phi} \delta_{\mu \nu},
\label{confgauge}
\ee
which requires us to introduce the bosonic (commuting) ghosts $\beta$ and
$\gamma$ in addition to the fermionic (anticommuting) ghosts $b$ and $c$. The
ghosts $b$ and $\beta$ have spin two, whereas $c$ and $\gamma$ have spin $-1$.
Of course, the central charge of this ghost multiplet is again $c=0$, by our
familiar theme of `bosons cancel fermions.' It is an easy exercise to write
down the generators $T$, $G$, $Q$, and $J$ for the ghost system, and verify
that they obey the relations we have come to expect.

In terms of the conformal factors $\phi$ and $\rho$ the $Q$-transformation of
the metric reads
\be
\delta \phi = \rho,\qquad \delta \rho = 0.
\label{confmode}
\ee
Since the total central charge vanishes, we can fix the Weyl transformations
that shift these modes by the conditions $\d \dbar \phi = \Rhat$ and $\d \dbar
\rho =0$, where $\Rhat$ is the world-sheet curvature of a fiducial background
metric $\hat{g}_{\mu\nu}$. These constraints will be implemented with the aid
of Lagrange multipliers $\pi$ and $\lambda$. Taking all together, the total
action for the theory is $$ S = S_{m}+S_{gh}+S_{L} , $$
\be
S_{gh} = \int d^2\!z\;( b \dbar c + \beta \dbar \gamma + {\rm c.c.}),
\ee
$$
S_{L} = \int d^2\!z\;(\pi (\d\dbar \phi - \Rhat) - \lambda \d \dbar \rho).
$$
In order to find the physical spectrum we have to know the correct BRST charge.
In fact, here we have two candidates. First, we have a topological field theory
that comes naturally with a nilpotent $Q$ charge---we have been very careful to
preserve this feature. Secondly, we are dealing with a gauge theory complete
with ghosts and its own BRST charge, since we gauged the symmetries generated
by the stress tensor $T$ and its fermionic partner $G$. This BRST charge is of
the form
\be
Q_{BRST}
= \oint {dz \over 2\pi i}
:\! c\left( T_m +T_L +\half T_{gh} \right) + \gamma \left( G_m +G_L
+\half G_{gh} \right)\! :
\label{Qbrstmm}
\ee
and is a simple generalization of the well-known $Q_{BRST}$ of bosonic string
theory. A precise analysis shows that in this case the correct BRST charge,
whose cohomology determines the spectrum of the theory, is the sum of these two
charges
\be
Q_{total} = Q + Q_{BRST}
\ee
An analysis, that we will not repeat here, shows that the physical operators
with respect to $Q_{total}$ are of the form
\be
\cO = \phi_i \cdot \st_n,
\ee
where $\phi_i$ is a matter field with charge $q_i$ and $\st_n$ is a
combination of ghost and Liouville fields with charge $n$, a non-negative
integer. These gravitational fields $\st_n$ are a new feature of topological
string theory---they will of course correspond to the Mumford-Morita-Miller
classes of \S2.

We now must prescribe how to calculate correlation functions of these
operators. As in critical string theory, we define the higher genus correlation
functions so that they become volume forms on the moduli space of Riemann
surfaces. These forms are then integrated to get the final correlation
functions, using the Beltrami differentials $\mu_1,\ldots,\mu_{3g-3}$. The
precise definition is
\be
\l {\cal O}_1 \ldots {\cal O}_n \r\subg \equiv
\Int_{{\cal M}_g} \l \int {\cal O}_1^{(2)}
\ldots \int {\cal O}_n^{(2)} \prod ^{3g-3}_{a=1} \int
\mu_a G \, \int \mubar_a \Gbar \, \r\subg.
\label{tstcorr}
\ee
The operators are of the form $\OO_k = \phi_{i_k} \cdot \st_{n_k}$, and
satisfy the selection rule
\be
\sum_k (n_k + q_{i_k} - 1)=(d-3)(1-g).
\label{TSTsrule}
\ee
The total background charge on the RHS is composed of the matter contribution
$d(1-g)$ and a term needed to cancel the charge of the $3g-3$ $G$-insertions.
The correlation function (\ref{tstcorr}) is a generalization of the one for
critical string theory, with the anti-ghosts $b(z)$ are replaced by the
supercurrents $G(z)$, see \cite{notes} for further comments on this analogy.

\newsubsection{Pure Topological Gravity}

We will proceed to solve topological string theory in the case of pure
topological gravity. That is, the case that no matter sector is present
($d=0$). Since we work with a cohomological theory, there are many possible
representations of the operators. We will follow here \cite{verlinde} and
choose a representation in which the correlation functions receive no
contribution from the interior of moduli space.

Let us start with the definition of the operators $\st_n$
\be
\st_n = 3^n e^{{2\over3}(n-1)\pi}(\gammatil)^n,~~~~~{\rm with}~~~~
\gammatil = \dbar \left( \half \d c + c\d \phi - \cc \right).
\ee
In order to understand this definition a bit more intuitively, introduce
$Q=Q_0+\Qbar_0$ and $Q_-=Q_0-\Qbar_0$. Since these operators have an
interpretation as Dolbeault operators $\d$, $\dbar$ on the world-sheet, one can
verify that
\be
\gammatil = \{ Q,\{Q_-,\phi\}\} \sim \d \dbar \phi = \Rhat
\label{gammacurv}
\ee
So $\gammatil$ is roughly the curvature, and $\st_n$ corresponds to the $n\th$
power of $\Rhat$. A much precise analysis \cite{verlinde,notes} shows that, in
the notation of \S2, the operator $\st_n$ is related to the cohomology class
$\sigma_n \in H^*(\Mgs)$ by
\be
\st_n = (2n+1)!! \cdot \sigma_n = \cO_{2n+1}.
\ee
We note that $\gammatil$ naively appears to decouple because it is BRST exact.
However, the model is not well-defined as it stands. To define the correlation
functions unambiguously, the fields must be restricted to a subspace $\HH_0$ of
the `big' Hilbert space $\HH'$
\be
\HH_0 \equiv \left\{ |\phi \rangle \in \HH' \ ; \
(b_0-\bbar_0)|\phi \rangle = 0 \right\} ;
\ee
otherwise, there is an obstruction to globally define correlation function over
$\Mgs$ \cite{j-p}. So the physical Hilbert space $\HH=H^*_Q(\HH_0)$ is
composed of $Q$ cohomology classes equivariant with respect to $b_0-\bbar_0$.
This insures that local coordinates can be consistently defined at the
punctures. It can easily be seen that, even though $\gammatil$ is BRST exact in
$\HH'$, it is not in $\HH_0$ and therefore does not decouple.

\newsubsection{Contact and factorization terms}

As was shown in \cite{verlinde} the correlation functions for pure topological
gravity are completely determined from the boundary of moduli space.
In fact, contact terms, which occur whenever two operators collide and are
quite generally of the form
\be
\cO_i(z) \cdot \cO_j(0) \sim \cO_k(0) \delta^2(z),
\ee
suffice to solve the model. To see this we must work out the contact term
algebra explicitly. We will sketch the argument of \cite{verlinde}.

Define the states $|\st_m \rangle = \st_m(0) | 0 \rangle$, and consider the
contribution of the operator $\st_n$ when it is close to the operator $\st_m$,
located at $z=0$ in some coordinates
\be
\int d^2\!z \cdot \st_n(z,\zbar)\cdot \st_m(0) |0\r =
\Int_{|q| < \epsilon} \frac{d^2q}{|q|^2} \, G_0 \,\Gbar_0\,
q^{L_0}\qbar^{\Lbar_0} \st_n (1) |\st_m\rangle.
\ee
Now recall that $\st_n\ \sim \tilde{\gamma}^n$ with $\tilde{\gamma}=\{ Q, \{
Q_-, \phi \} \}$. Since $L_0$ is $Q$ exact, we find a contribution due to
\be
\int d^2\!q \; \d_q \d_\qbar \,q^{L_0} \qbar^{\Lbar_0} \phi (0) | \st
_{n+m-1} \rangle,
\label{int2}
\ee
where $q^{L_0} \qbar^{\Lbar_0} \phi (0) \sim \log |q|^2 + {\rm regular}$,
because of the anomalous covariance of the field $\phi$. This clearly gives a
$\delta$-function contribution, and with the inclusion of all factors one finds
in the end
\be
\int d^2\!z\,\st_n(z,\zbar) \st_m(0) = (2m+1) \st_{n+m-1}(0).
\label{sigs}
\ee
Note that this expression is not symmetric in $n$ and $m$,
\be
\int \left[ \st_n, \st_m \right] = 2(m-n) \st_{m+n-1},
\label{cont}
\ee
where we see the appearance of a Virasoro algebra.

Let us now consider a general correlator of the form
\be
\l \st_n \st_{k_1}\ldots \st_{k_s} \r\subg.
\ee
We want to study the contributions to this amplitude that we obtain if we
integrate the two-form $\st_n$ over the Riemann surface. In general there are
three different contributions: the contact terms that we examined above,
factorization terms, and, finally, bulk terms. The factorization terms arise
when the surface itself degenerates---the cases {\it (ii)} and {\it (iii)} of
\fig2. They may be expressed as sums over factorized correlation functions with
$\st_i$ insertions, and are, by charge conservation, of the general form
\be
\sum_{i+j=n-2} A^{(n)}_{ij}
\l \st_i \st_{k_1} \ldots \st_{k_r} \r_{\strut g'}
\l \st_j \st_{k_{r+1}} \ldots \st_{k_s} \r_{\strut g-g'}
\ee
or
\be
\sum_{i+j=n-2} B^{(n)}_{ij}
\l \st_i \st_j \st_{k_1} \ldots \st_{k_s} \r_{\strut g-1}
\ee
with some coefficients $A^{(n)}_{ij},B^{(n)}_{ij}$. The bulk terms describe the
contributions from the interior of moduli space. Quite remarkably, one can show
that these contributions are absent due to the asymmetry in the contact terms.
Also the factorization terms are completely determined by consistency:
$A^{(n)}_{ij}=B^{(n)}_{ij}=\half$. So, in the end we are left with the
following recursion relations for the correlation functions of topological
gravity.
\ba
\l \st_n \prod_{k \in S} \st_k \r\subg \iss
\sum_{k \in S} (2k+1) \l \st_{n+k-1} \prod_{k\p \neq k}
\st_{k\p} \r\subg +
\sum_{i+j=n-2} \half \l \st_i \st_j \prod \st_k \r_{\strut{g-1}} \nonu
& & \qquad + \sum \ppinch \half
\l \st_i \prod_{k \in X} \st_k \r_{\strut g'}\,
\l \st_j \prod_{k \in Y} \st_k \r_{\strut g-g'}.
\label{recurs}
\ea
Here $S=\{k_1,\ldots,k_s\}$, and the last summation is over all partitions of
$S$ into two subsets $X,Y$. The three terms on the RHS are associated to the
three inequivalent degenerations of \fig2: contact terms and factorization
terms of a homology cycle or dividing cycle. The contributions always come from
the boundary of moduli space, so the recursion relation lowers the genus $g$
and/or the number of operators $s$, and ultimately gives factors of the
three-point
function on the sphere,
\be
\l\st_0^3\r\subzero = \l\sigma_0^3\r\subzero= 1,
\ee
or the one-point function on the torus
\be
\l\st_1\r\subone = 3\l\sigma_1\r\subone = {1\over8}.
\ee
Therefore the recursion relations suffice to calculate any intersection number.
In the case $n=0$ the relation (\ref{recurs}) is nothing but the puncture
equation (\ref{puncture}); the case $n=1$ is a linear combination of the
dilaton equation (\ref{dilaton}) and the ghost charge conservation
(\ref{charge}). Note that in these two cases the factorization terms are
absent.

Although these relations are manifestly non-linear in terms of the free energy
$F$, they turn out to be linear when written in terms of the partition function
$\tau=e^{F}$, due to the wonderful relation
\be
\tau\inv{\d^2\tau}= \d^2 F + (\d F)^2.
\ee
Indeed, consider the partition function
\be
\tau(\tt) = \exp \sum_{g=0}^\infty \Bl \exp \sum_n
\tt_n \st_n \Br\subg
\ee
where the coupling coefficients $\tt_n$ are related to the KdV times by
$\tt_n=t_{2n+1}$. Now, if we recall that a derivative $\d \tau/\d\tt_n$
corresponds to an insertion of $\st_n$, whereas a multiplication $\tt_n\cdot
\tau$ removes the same operator from the generating function, the recursion
relations (\ref{recurs}) can be expressed as linear, homogeneous differential
equations for the $\tau$-function
\be
L_n \cdot \tau=0,
\qquad (n\geq -1).
\label{vir}
\ee
where $L_n$ denote the differential operators
\ba
L_{-1} \iss - \half \ddtt 0 +
\sum_{k=1}^\infty (k+\half) \tt_k \ddtt{k-1} + \fr14 \tt_0^2\nonu
\label{vir2}
L_0 \iss - \half \ddtt1 + \sum_{k=0}^\infty (k+\half)\tt_k \ddtt k +
\fr1{16}\qquad,
\\[.5mm]
L_n \iss - \half \ddtt {n-1} +\sum_{k=0}^\infty (k+\half)
\tt_k \ddtt{k+n} + \fr14
\sum_{i=1}^{n}{\partial^2 \over \partial \tt_{i-1} \partial \tt_{n-i}}.
\nonumber
\ea
These are exactly the equations that according to \S4.4 characterize the
$\tau$-function of the KdV hierarchy that we studied in such great detail
before. So, the above manipulations amount to a sketch of a
quantum field theory proof of Witten's original conjecture.

In the more general case of a topological minimal model coupled to gravity,
these Virasoro constraints are replaced by the $W$-constraints
\be
W^{(s)}_n \cdot \tau= 0,\qquad s=2,\ldots,p,\ n \geq 1-s,
\ee
The differential operators $W^{(s)}_n$ form a $W$-algebra associated to the Lie
group $G$ that labeled the minimal model. They are spin $s$ currents that
generalize the Virasoro algebra. We have one $W$-generator for each Casimir in
the group $G$. The constraints now obtain an interpretation as multiple contact
and factorization terms \cite{loop,Li}. A full discussion of these relations is
however completely outside the scope of these lectures.

\vspace{.8cm}

\noindent{\bf Acknowledgements}

I wish to thank the organizers of this wonderful school for inviting me to
present these lectures and for creating such a stimulating and enjoyable
atmosphere. Needless to say I benefitted greatly from countless insightful
discussion with my collaborators Erik Verlinde, Herman Verlinde, and Edward
Witten. I further thank Rob Rudd for his assistance in preparing a preliminary
version of these notes.

\renewcommand{\Large}{\normalsize} 
\end{document}